\documentclass[11pt, epsf]{amsart}
\usepackage{epsfig}
\usepackage{eucal}
\usepackage{subfig}
\usepackage{amssymb}
\usepackage{latexsym}
\usepackage{amsfonts}
\usepackage[latin1]{inputenc}
\usepackage{slashbox}
\usepackage{colortbl}
\usepackage{float}

\newenvironment{namelist}[1]{%
\begin{list}{}
 {
   
   \settowidth{\labelwidth}{#1}
   \setlength{\leftmargin}{1.1\labelwidth}
  }
 }{%
\end{list}}
\newcommand{\bi}{\begin{namelist}}
\newcommand{\ei}{\end{namelist}}

\newcommand{\df}{\displaystyle}
\newcommand{\eps}{\epsilon}

\newtheorem{Def}{Definition}

\newtheorem{Theo}{Theorem}

\newtheorem{Lem}{Lemma}

\title{$hp$-Cloud Approximation Of The Dirac Eigenvalue Problem: The Way Of Stability}
\author{Hasan Almanasreh}
\thanks{Department of Mathematical Sciences, University of Gothenburg,
        SE-412 96 Gothenburg, Sweden}

\begin{document}

\maketitle

\begin{abstract} We apply $hp$-cloud method to the radial Dirac eigenvalue problem. The difficulty of occurrence of spurious eigenvalues among the genuine ones in the computation is resolved. The method of treatment is based on assuming $hp$-cloud Petrov-Galerkin scheme to construct the weak formulation of the problem which adds a consistent diffusivity to the variational formulation. The size of the artificially added diffusion term is controlled by a stability parameter ($\tau$). The derivation of $\tau$ assumes the limit behavior of the eigenvalues at infinity. The parameter $\tau$ is applicable for generic basis functions. This is combined with the choice of appropriate intrinsic enrichments in the construction of the cloud shape functions.\\

\end{abstract}

\section{Introduction.}

In the last decades, several numerical methods have been derived to compute the eigenvalues of operators. The need of accurate computation of eigenvalues is considered due to their significant applications in many disciplines of science. Mathematically, if a matrix or a linear operator is diagonalized, then by the spectral theorem, it can be analyzed by studying its corresponding eigenvalues, i.e., transforming the matrix or operator to a set of eigenfunctions which can be easily studied. From the physical point of view, the eigenvalues possess a wide range of information about the behavior of the system governed by an operator. This information might be all what is needed to answer many questions regarding the system properties such as stability, positivity, boundedness, asymptotic behavior, etc. In mechanics, eigenvalues play a central role in several aspects such as determining whether the automobile is noisy, whether a bridge will collapse by the water waves, etc. Also, the eigenvalues describe how the quantum state of a physical system changes in time (Schr\"odinger equation). They also represent the electrons relativistic energies and describe their motion in the atomic levels, this is the well-known Dirac equation, which is the core of the present work.

The calculation of energy levels in Helium-like ions, where the interaction between two electrons takes place, can be determined by studying the electrons correlation which is part of quantum electrodynamic effects (QED-effects). A scheme for calculating QED-effects \cite{LIND,MOH,ROS,SHAB1} is based on a basis set of relativistic Hydrogen-like ion wave eigenfunctions (of the Dirac operator). Meanwhile, the numerical computation of the Dirac operator eigenvalues encounters unphysical values (do not match the physical observations) called spurious eigenvalues or spectrum pollution. The spurious eigenvalues result in rapid oscillations in the wave functions, hence, in many cases, affecting the computation reliability of the basis set in the practical atomic calculations.

The spurious eigenvalues are an effect of the numerical methods and are found in the computation of many problems other than the Dirac eigenvalue problem \cite{ACK,ALMA,PES,SHAB}. For general eigenvalue problems, spurious eigenvalues are reported in \cite{ZHA}. The occurrence of the spuriosity is related to mismatching of desired properties of the original solution in the numerical formulation. Also in the computation of electromagnetic problems the spuriosity is perceived \cite{MUR,SCH}. Two leading approaches are derived to solve this difficulty; Shabaev et al. \cite{SHAB} have related the spuriosity to the symmetric treatment of the large and small components of the Dirac wave function. Their approach, for removing the spurious eigenvalues, is based on deriving dual kinetic-balance (DKB) basis functions for the large and small components. Almanasreh et al. \cite{ALMA} have allied the occurrence of spurious eigenvalues to the incorrect treatment of the trial and test functions in the finite element method (FEM). They proposed a stability scheme based on creating diffusivity by modifying the test function so that it includes a gradient-based correction term.

In this work, we apply $hp$-cloud method \cite{DUA,ZUP} to the radial Dirac eigenvalue problem. The technique is known as one of the meshfree methods (MMs) \cite{BEL2,FRI2,LIU,LU,NGU} that allows two different enrichments, intrinsic and extrinsic, to be built in the construction of the basis functions. The method was previously applied for several problems, e.g., the Schr\"odinger equation \cite{CHE2}, Mindlin's thick plate model \cite{GAR}, and Timoshenko beam problems \cite{MEN}, etc. Here, we apply $hp$-cloud method based on the Galerkin formulation. This means that it is required to evaluate the integrals in the weak formulation of the particular equation, thus a background mesh must be employed in the integration techniques. Therefore, the $hp$-cloud method used here is not really a truly MM. However, all other features of MMs are maintained in our approximation.

In order to treat the spuriosity problem, we stabilize the computation by considering instead an $hp$-cloud Petrov-Galerkin ($hp$-CPG) method which may be considered as a new technique of the general meshfree local Petrov-Galerkin (MLPG) \cite{ATL,FRI1,LIN}. The stability scheme is based on adding consistent diffusion terms without changing the structure of the equation. The size of the additional diffusivity is controlled by a stability parameter.

Consider the radial Dirac eigenvalue problem $H_\kappa\Phi(x)=\lambda\Phi(x)$, where $\Phi(x)=(F(x),G(x))^t$ is the radial wave function, $\lambda$ is the electron relativistic energies (eigenvalues), and $H_\kappa$ is the radial Dirac operator given by
\vspace{-0.1cm}
\begin{equation*}
H_\kappa = \left(
\begin{array}{cc}
mc^2+V(x) & c\big(\!-\!D_x+\frac{\kappa}{x}\big) \\
c\big(D_x+\frac{\kappa}{x}\big) & -mc^2+V(x)
\end{array}
\right)\,.
\vspace{-0.1cm}
\end{equation*}
The constant $c$ is the speed of light, $m$ is the electron mass, $V$ is the Coulomb potential, $D_x=d/dx$, and $\kappa$ is the spin-orbit coupling parameter defined as $\kappa=(-1)^{\jmath+\ell+\frac{1}{2}}(\jmath+\frac{1}{2})$, where $\jmath$ and $\ell$ are the total and the orbital angular momentum quantum numbers respectively. The weak formulation of the problem is to find $\lambda\in\mathbb{R}$ and $\Phi$ in a specific function space such that for every test function $\Psi$ in some suitable space we have $\int_{\Omega}\Psi^t H_\kappa \Phi dx=\lambda\int_{\Omega}\Psi^t \Phi dx$. The usual $hp$-cloud Galerkin approximation is to let $\Psi$ to be $(\psi,0)^t$ and $(0,\psi)^t$, where $\psi$ is in the same space as of the two components of $\Phi$. To discretize the weak form, the components of the trial function $\Phi$ and the test function $\psi$ are chosen from a finite set of $hp$-cloud basis functions which are constructed using moving least-squares method. Since the radial Dirac operator is dominated by advection (convection) terms, the $hp$-cloud approximation will be upset by spurious eigenvalues.

To stabilize the $hp$-cloud approximation, the $hp$-CPG method is used. In this formulation, the test function $\Psi$ is assumed to belong to a function space different from that of the trial function $\Phi$, in the sense that $\Psi$ is chosen in the form $(\psi,\tau\psi')^t$ and $(\tau\psi',\psi)^t$ where $\psi$ belongs to the same space as the two components of $\Phi$. The correction term $\tau\psi'$ is used to add artificial viscosity, controlled by $\tau$, to stabilize the convection terms. The derivation of the stability parameter $\tau$ follows the principle used in \cite{ALMA} for the FEM. Two assumptions are considered in deriving $\tau$; $(i)$ the operator limit as the radial variable $x$ tends to infinity, thus obtaining an approximation of the limit point of the eigenvalues (depending on $\tau$) which can be compared to the theoretical limit point eigenvalue \cite{GRI}, $(ii)$ considering the dominant terms with respect to the speed of light ($c$).\\

The paper is organized as follows; in Section 2, some preliminaries about the Dirac equation are presented, also we shed some light over the occurrence of the spuriosity. In Section 3, the construction of the $hp$-cloud functions is provided, also coupling with the FEM to impose essential boundary conditions (EBCs) is explained. The $hp$-CPG method and the derivation of the stability parameter are treated in Section 4. In the last section, Section 5, we present some numerical results and provide necessary discussion about the stability scheme.
\section{The radial Dirac eigenvalue problem and the spuriosity}
The free Dirac space-time equation is given by
\begin{equation}\label{1}
i\,\hslash\frac{\partial}{\partial t}\mathbf{u}(\textbf{x},t) =\mathbf{H}_0\mathbf{u}(\textbf{x},t)\, ,\quad\quad \mathbf{u}(\textbf{x},0)=\mathbf{u}_0(\textbf{x}),
\end{equation}
where $\hslash$ is the Planck constant divided by $2\pi$, and $\mathbf{H}_0:  H^1({\mathbb R}^3;{\mathbb C}^4)\longrightarrow  L^2({\mathbb R}^3;{\mathbb C}^4)$ is the free Dirac operator acting on the four-component vector $\mathbf{u}$, given by
\begin{equation}\label{2}
\mathbf{H}_0 = -i\,{\hslash}c\boldsymbol{\alpha}\cdot \mathbf{\nabla} + mc^2 \beta\,.
\end{equation}
The $4\times 4$ Dirac matrices $\boldsymbol{\alpha}=(\alpha_1,\alpha_2,\alpha_3)$ and $\beta$ are  given by
$$
\alpha_j = \left(
\begin{array}{cc}
0 & \sigma_j \\
\sigma_j & 0
\end{array}
\right)\;\;\text{and}\;\;
\beta = \left(
\begin{array}{cc}
I & 0 \\
0 & -I
\end{array}
\right)\,.
$$
Here $I$ and $0$ are the $2\times 2$ unity and zeros matrices respectively, and $\sigma_j$'s are the $2\times 2$ Pauli matrices
$$
\sigma_1 = \left(
\begin{array}{cc}
0 & 1 \\
1 & 0
\end{array}
\right),\;\;
\sigma_2 = \left(
\begin{array}{cc}
0 & -i \\
i & 0
\end{array}
\right)\, ,
\;\;\text{and}\;\;
\sigma_3 = \left(
\begin{array}{cc}
1 & 0 \\
0 & -1
\end{array}
\right)\, .
$$
Note that separation of variable yields the Dirac eigenvalue problem, i.e., by assuming $\mathbf{u}(\textbf{x},t)=u(\textbf{x})\theta(t)$ in (\ref{1}) one gets
\begin{equation}\label{3}
\mathbf{H}_0u(\textbf{x})=\lambda u(\textbf{x}).
\end{equation}
The operator $\mathbf{H}_0$ is self adjoint on $ H^1({\mathbb R}^3;{\mathbb C}^4)$, it describes the motion of the electron that moves freely without external forces. The free Dirac operator with Coulomb potential is given as
\begin{equation}\label{4}
\mathbf{H}=\mathbf{H}_0+V(\textbf{x})I\, ,
\end{equation}
where $V(\textbf{x})=\frac{-Z}{|\textbf{x}|}$, and $I$ is the $4\times 4$ unity matrix. The spectrum, denoted by $\sigma$, of the Coulomb-Dirac operator is $\sigma(\mathbf{H}) = (-\infty,-mc^2]\cup\{\lambda_k\}_{k\in\mathbb N}\cup[mc^2,+\infty)$, where $\{\lambda_k\}_{k\in\mathbb N}$ is a discrete sequence of eigenvalues in the gap $(-mc^2,mc^2)$ of the continuous spectrum.

The radial Coulomb-Dirac operator (radial Dirac operator) can be obtained by separation of variables of the radial and angular parts, i.e., by assuming
$u(\textbf{x})=\displaystyle\frac{1}{x}\left(
\begin{array}{c}
F(x)\chi_{\kappa,m}(\varpi,\Theta) \\
i\,G(x)\chi_{-\kappa,m}(\varpi,\Theta)
\end{array}
\right)$, where $x$ represents the radial variable. The spherical symmetry property of the angular function $\chi$ is the key point in the derivation of the radial part. Let $\Phi(x)=(F(x),G(x))^t$, the radial Dirac eigenvalue problem is given as
\begin{equation}\label{5}
H_\kappa\Phi(x)=\lambda\Phi(x),\; \text{where}\;
\end{equation}
\begin{equation}\label{6}
H_\kappa = \left(
\begin{array}{cc}
mc^2+V(x) & c\big(\!-\!D_x+\frac{\kappa}{x}\big) \\
c\big(D_x+\frac{\kappa}{x}\big) & -mc^2+V(x)
\end{array}
\right)\,.
\end{equation}
The radial functions $F(x)$ and $G(x)$ are called respectively the large and small components of the wave function $\Phi(x)$.

The well-known difficulty of solving the radial Dirac eigenvalue problem numerically is the presence of spurious eigenvalues among the genuine ones that disturb the solution and consequently affect the reliability of the approximated eigenstates. Here we follow \cite{ALMA} for the classification of the spurious eigenvalues; the first category is the so-called instilled spuriosity, and the second category is the unphysical coincidence phenomenon. The first type is those spurious eigenvalues that may occur within the true eigenvalues (i.e., they occur between the genuine energy levels), this type of spuriosity appears for all values of $\kappa$. The second type is the unphysical assigning of almost the same first eigenvalue for $2s_{1/2}(\kappa=-1)$ and $2p_{1/2}(\kappa=1)$, $3p_{3/2}(\kappa=-2)$ and $3d_{3/2}(\kappa=2)$, $4d_{5/2}(\kappa=-3)$ and $4f_{5/2}(\kappa=3)$, and so on. This phenomenon is rigorously studied in \cite{THA} from theoretical aspect and in \cite{TUPS} from numerical viewpoints.

Apparently, most authors \cite{ACK,ALMA,PES,SHAB} agree that the incorrect balancing and the symmetric treatment of the large and small components of the wave function or of the test and trial functions in the numerical methods are the core of the problem. In the present work, we relate the occurrence of spuriosity, of both categories, to the unsuitable function spaces and to the symmetric treatment of the trial and test functions in the weak formulation of the equation. To clarify, we rewrite (\ref{5}) to obtain an explicit formula for each of the two radial functions, so by defining $w^{\pm}(x)=\pm mc^2+V(x)$ we have, see \cite{ALMA},
\begin{equation}\label{8}
F''(x)+\gamma_1(x,\lambda)F'(x)+\gamma_2(x,\lambda)F(x)=0\, ,
\end{equation}
\begin{equation}\label{9}
G''(x)+\theta_1(x,\lambda)G'(x)+\theta_2(x,\lambda)G(x)=0\, ,
\end{equation}
where
$$
\gamma_1(x,\lambda)=-\frac{V'(x)}{w^{-}(x)-\lambda}\, ,\quad \theta_1(x,\lambda)=-\frac{V'(x)}{w^{+}(x)-\lambda}\, ,
$$
$$
\gamma_2(x,\lambda)=\frac{\big(w^{+}(x)-\lambda\big)\big(w^{-}(x)- \lambda\big)}{c^2}-\frac{\kappa^2+
\kappa}{x^2}-\frac{\kappa V'(x)}{x\big(w^{-}(x)-\lambda\big)}\, ,
$$
and
$$
\theta_2(x,\lambda)=\frac{\big(w^{+}(x)-\lambda\big)\big(w^{-}(x)- \lambda\big)}{c^2}-\frac{\kappa^2-
\kappa}{x^2}+\frac{\kappa V'(x)}{x\big(w^{+}(x)-\lambda\big)}\, .
$$

According to (\ref{8}) and (\ref{9}), the components $F$ and $G$ should be continuous and have continuous first derivatives. Thus, the suitable choices of function spaces for the computation of the radial Dirac eigenvalue problem are those that possessing these properties. Assuming appropriate spaces helps in partial elimination of spurious eigenvalues, and does not help overcoming the unphysical coincidence phenomenon. In \cite{ALMA}, the same argument is accounted, where the FEM is applied to approximate the eigenvalues using linear basis functions.
\begin{table}[h]
\begin{footnotesize}
\caption{This table, taken from \cite{ALMA}, shows the first computed eigenvalues of the electron in the Hydrogen atom.}
\centering
\begin{tabular}{@{} l c c r @{}}
\hline\hline
Level & $\kappa=1$ & $\kappa=-1$ & Rel. Form. $\kappa=-1$ \\ [0.5ex]
\hline
1 &\cellcolor[gray]{0.6} -0.50000665661 & -0.50000665659 & -0.50000665659 \\
2 & -0.12500208841 & -0.12500208839 & -0.12500208018 \\
3 & -0.05555631532 & -0.05555631532 & -0.05555629517 \\
 $\Rrightarrow$    &\cellcolor[gray]{0.6} -0.03141172061 &\cellcolor[gray]{0.6} -0.03141172060 & Spurious Eigenvalue \\
4 & -0.03118772526 & -0.03118772524 & -0.03125033803 \\
5 & -0.01974434510 & -0.01974434508 & -0.02000018105 \\ [1ex]
\hline\hline
\end{tabular}
\label{table:nonlin}
\end{footnotesize}
\end{table}

In Table 1, 400 nodal points are used to discretize the domain, and the computation is performed for point nucleus. The shaded value in the first level is what meant by the unphysical coincidence phenomenon, and the two shaded values after the third level are the so-called instilled spuriosity. If the basis functions are chosen to be $C^1$-functions, then some instilled spurious eigenvalues are avoided as indicated in \cite{ALMA}. Therefore, after applying the boundary conditions, homogeneous Dirichlet boundary condition is assumed for both radial functions, the proposed space is $\mathcal{H}(\Omega):=C^1(\Omega)\cap H^1_0(\Omega)$. Also, the radial functions are mostly like to vanish at the boundaries in a damping way (except some states), consequently homogeneous Neumann boundary condition should be taken into account. The exceptional states are $1s_{1/2}$ and $2p_{1/2}$, in this case at the upper boundary the same treatment is considered as of the other states, but the first derivative of these two states at the lower boundary is not zero. Here we will assume general boundary condition, that is, homogeneous Dirichlet boundary condition. Thus, from now on, the space $\mathcal{H}(\Omega)$ is considered.

What deserves to dwell upon is that in numerical methods applied to convection dominated problem, the solution is disturbed by spurious oscillations. This instability gets worse as the convection size increases. The following two numbers are considered as tools to measure the dominance of the convection term
\begin{equation}\label{10}
Pe_j=\frac{|u_j|h_j}{2K}\;\; \text{and}\;\;
Da_j=\frac{s_j h_j}{|u_j|}\, ,
\end{equation}
where $Pe_j$ and $Da_j$ are known as the grid Peclet and Damk\"ohler numbers respectively, $h_j$ is the size of the element interval $I_j$, $u_j$ and $s_j$ are respectively the coefficients of the convection and the reaction terms corresponding to $I_j$, and $K$ is the diffusivity size. The difficulty arises when either the convection coefficient or the source term is larger than the diffusion coefficient, i.e., when $Pe_j>1$ or when $2Pe_jDa_j=\df\frac{s_jh_j^2}{K}>1$ respectively, then the associated equation is a convection dominated one.

For the simplified equations (\ref{8}) and (\ref{9}), it is easy to check that $2PeDa$ admits very large values if small number of nodal points is considered regardless the sizes of $|\lambda|$, $Z$, and $\kappa$. Even with mesh refinement, $2PeDa$ still admits very large values at some positions ($2Pe_jDa_j>>1$ for some $j$). The Peclet number, $Pe$, for the equation that involves the function $F$, is always less than 1. But for the equation that corresponds to $G$, $Pe$ admits a value greater than one, exactly at the nodal point where $u_j$ changes its sign, here refining the mesh does not bring $Pe$ below one for all choices of $\lambda$, $Z$, and $\kappa$. Hence, (\ref{8}) and (\ref{9}) are dominated by convection terms. Thus the approximated solutions $F$ and $G$, will be upset by unphysical oscillations. The draw back is that these oscillations in the eigenfunctions result in some unphysical eigenvalues, the spurious eigenvalues. For detailed study on convection dominated problems we refer to \cite{BRO81,BRO82}.

\section{Moving least-squares (MLS) approximation}
To build the $hp$-cloud functions, MLS method is applied which allows easily $p$-enrichment to be implemented and to desired fundamental characters to be enriched in the approximation. MLS is a well-known approximation technique for constructing meshfree shape functions in general. It applies certain least square approach to get the best local approximation, then the local variable is let to 'move' to cover the whole domain.\\
Consider an open bounded domain $\Omega\subset\mathbb{R}$ with boundary $\partial \Omega$, assume $X=\{x_1,x_2,\ldots, x_n\}$ is a set of $n$ arbitrary points in $\overline{\Omega}$. Let $W=\{w_i\}_{i=1}^n$ be a set of open covering of $\Omega$ defined by $X$ such that $w_i$ is centered at $x_i$ and $\Omega\subset\cup_{i=1}^n w_i$.
\begin{Def}
A set of functions $\{\psi_i\}_{i=1}^n$ is called a partition of unity (PU) subordinated to the cover $W$ if
\begin{itemize}
\item [(1)] $\sum_{i=1}^n\psi_i(x)=1$ for all $x\in\Omega$.
\item [(2)] $\psi\in C_0^s(w_i)$ for $i=1,2,\ldots, n$, where $s\geq0$.
\end{itemize}
\end{Def}

Let $P=\{p_1(x), p_2(x), \ldots, p_m(x)\}$ be a set of basis of polynomials (or any basis of suitable intrinsic enrichments). Suppose that $\Psi(x)$ is a continuous function defined on $\Omega$ and that its values, $\Psi_i$, at the points $x_i\in\overline{\Omega}$, $i=1, 2, \ldots, n$, are given. To approximate $\Psi(x)$ globally by $\Psi_h(x)$, firstly $\Psi(x)$ is approximated locally at $\tilde{x}\in\overline{\Omega}$ by $J_{\tilde{x}}\Psi$ defined in terms of the given basis set $P$ as
\begin{equation}\label{30}
J_{\tilde{x}}\Psi(x)=P^t(x)a(\tilde{x}),
\end{equation}
where $t$ denotes the usual transpose. The unknown coefficients $a(\tilde{x})$ are chosen so that $J_{\tilde{x}}\Psi$ is the best approximation of $\Psi$ in a certain least squares sense, this is achieved if $a$ is selected to minimize the following weighted least squares $L^2$-error
\begin{equation}\label{31}
I_{\tilde{x}}(a)=\displaystyle \sum_{i=1}^n \varphi_i (\frac{x-x_i}{\rho_i})(P^t(x_i)a(\tilde{x})-\Psi_i)^2,
\end{equation}
where $\varphi_i$ is a weight function introduced to insure the locality of the approximation, and $\rho_i$ is the dilation parameter which controls the support radius of $\varphi_i$ at $x_i$. To minimize $I_{\tilde{x}}$ with respect to the vector $a$, the first derivative test is applied, i.e., we set $\frac{\partial I_{\tilde{x}}}{\partial a}=0$ which gives
\begin{equation*}
\frac{\partial I_{\tilde{x}}}{\partial a_j}=\displaystyle \sum_{i=1}^n \varphi_i (\frac{x-x_i}{\rho_i})2p_j(x_i)(P^t(x_i)a(\tilde{x})-\Psi_i)=0,\quad j=1,2,\ldots,m.
\end{equation*}
The above system can be written as
\begin{equation}\label{32}
M(x)a(\tilde{x})-B(x)\Psi=0,
\end{equation}
where $M(x)=\mathcal{P}^t\Upsilon(x)\mathcal{P}$, $B(x)=\mathcal{P}^t\Upsilon(x)$, $\Psi^t=[\Psi_1,\Psi_2,\ldots, \Psi_n]$, and $a^t(\tilde{x})=[a_1(\tilde{x}),a_2(\tilde{x}),\ldots, $ $a_m(\tilde{x})]$, with $\mathcal{P}$ and $\Upsilon(x)$ defined as
$$
\mathcal{P}=\left(
\begin{array}{cccc}
p_1(x_1) & p_2(x_1) & \ldots & p_m(x_1) \\
p_1(x_2) & p_2(x_2) & \ldots & p_m(x_2) \\
\vdots   & \vdots   & \ddots   & \vdots \\
p_1(x_n) & p_2(x_n) & \ldots & p_m(x_n)
\end{array}
\right)
$$
and
$$
\Upsilon(x)=\left(
\begin{array}{cccc}
\varphi_1(\frac{x-x_1}{\rho_1}) & 0 & \ldots & 0 \\
0 & \varphi_2(\frac{x-x_2}{\rho_2}) & \ldots & 0 \\
\vdots   & \vdots   & \ddots   & \vdots \\
0 & 0 & \ldots & \varphi_n(\frac{x-x_n}{\rho_n})
\end{array}
\right).
$$
We proceed from equation (\ref{32}) to get
\begin{equation}\label{33}
a(\tilde{x})=M^{-1}(x)B(x)\Psi.
\end{equation}
Thus
\begin{equation*}
 J_{\tilde{x}}\Psi(x)=P^{t}(x)a(\tilde{x})
 =P^{t}(x)M^{-1}(x)B(x)\Psi.
\end{equation*}
The global approximations is then obtained by assuming $\tilde{x}$ arbitrary, i.e., by letting $\tilde{x}$ move over the domain, viz, the solution is globalized by considering $\Psi(x)\approx\displaystyle\lim_{\tilde{x}\to x}J_{\tilde{x}}\Psi(x)=:\Psi_h(x)$, thus
\begin{equation}\label{34}
\Psi_h(x)=\displaystyle\sum_{i=1}^n\psi_i(x)\Psi_i
\end{equation}
with $\psi_i(x)=P^t(x)M^{-1}(x)B_i(x)$, and $B_i(x)=\varphi_i(\frac{x-x_i}{\rho_i})P(x_i)$. To sum up, $\Psi_h$ can be written as
\begin{equation}\label{35}
\Psi_h(x)=P^t(x)\Big(\displaystyle\sum_{i=1}^n \varphi_i(\frac{x-x_i}{\rho_i})P(x_i)
P^t(x_i)\Big)^{-1}\displaystyle\sum_{i=1}^n \varphi_i(\frac{x-x_i}{\rho_i})P(x_i)\Psi_i.
\end{equation}
The first derivative of $\psi_i$ is given by $\psi_{i,x}=\frac{d\psi_i(x)}{dx}=P_x^tM^{-1}B_i-P^tM^{-1}M_xM^{-1}B_i+P^tM^{-1}B_{i,x}$. Below we shall need the consistency concept.
\begin{Def}
\emph{
A set of functions $\{u_i(x)\}$ is consistent of order $m$, if $\displaystyle\sum_{i}u_i(x)\mathfrak{P}(x_i)=\mathfrak{P}(x)$ for all $x\in\Omega$, where $\mathfrak{P}(x)=\{x^\varsigma;\;|\varsigma|\leq m\}$.
}
\end{Def}
To increase the order of consistency of the approximation, the complete representation of the $hp$-cloud functions consists of the set of PU functions $\psi_i(x)$ and monomial extrinsic enrichment basis functions $\mathrm{P}$ as
\begin{equation*}
\Psi_h(x)=\displaystyle \sum_{i=1}^n \psi_i(x) \Big(\sum_{j=1}^{n_0}\mathrm{P}_j(x)\Psi_i^j\Big)
 =\displaystyle \sum_{i=1}^n \sum_{j=1}^{n_0} \psi_i(x)\mathrm{P}_j(x)\Psi_i^j.
\end{equation*}
Note that $\mathrm{P}$ can be any type of basis functions, but the most used is monomials since they provide good approximation for smooth functions. The monomials $\mathrm{P}_j(x)$, according to \cite{ZUP}, should be normalized by the measure of the grid size at $x_j$ to prevent numerical instability. Nevertheless, in applying the $hp$-cloud approximation for the radial Dirac eigenvalue problem, we will use a stability scheme based on the MLPG method, for that we will not be interested in concerning extrinsic enrichments in the computation ($\mathrm{P}=\{1\}$, a monomial of degree zero). The point of this setting follows \cite{ATL}, where six different realizations of MLPG restricted only to intrinsic enrichment basis are considered. It is found that extrinsic enrichments in the MLPG method cause numerical stability problems, because the behavior of their derivatives has large oscillations, which is not the case in the usual MMs. Hence, in the present work, only intrinsic enrichments, $P(x)$, are considered, and thus the approximation with the $hp$-clouds is given by (\ref{34}).\\

The weight function $\varphi_i$ plays the most important role in the definition of the $hp$-cloud shape function, it is defined locally on the cover $w_i$ around $x_i$. The function $\varphi_i$ can also be chosen the same for all nodes, in this case we write $\varphi_i=\varphi$, which is the case we consider in this work. The $hp$-cloud, $\psi_i$, inherits the properties of the weight function $\varphi$ such as continuity and smoothness. In other words, if $\varphi$ is continuous with continuous first derivative, then so is $\psi_i$, provided that the continuity of the enrichment basis $P(x)$ and its first derivatives is ensured. As for the Dirac large and small components, $F$ and $G$, the proposed space is $\mathcal{H}$, therefore, the weight function $\varphi$ should be at least $C^1$-function. For this purpose, we will consider quartic spline (which is a $C^2$-function, sufficiently enough) as a weight function defined by
\begin{equation}\label{36}
\varphi(r)=\left\{
\begin{array}{ll}
1-6r^2+8r^3-3r^4\,,&r\leq1,\\
0\,,&r>1,
\end{array}
\right.
\end{equation}
where $r=\frac{|x-x_i|}{\rho_i}$.

The set functions $\{\psi_i\}_{i=1}^n$ builds a PU, also the set of their first derivatives $\{\psi_{i,x}\}_{i=1}^n$ builds a partition of nullity (PN) ($\sum_{i=1}^n\psi_{i,x}(x)=0$ for all $x\in\Omega$), see Figure 1. The computational effort of evaluating the integrals in the weak form of the $hp$-cloud approximation is more time consuming compared to mesh-based methods (the shape functions are of the form $\varphi_i$ only), this is due to the fact that the derivative of the shape function $\psi_i$ tends to have non-polynomial characters, also due to the time needed for matrix inversion in evaluating the shape functions.
\begin{figure}[h]
\centering
\includegraphics[width=7.8cm]{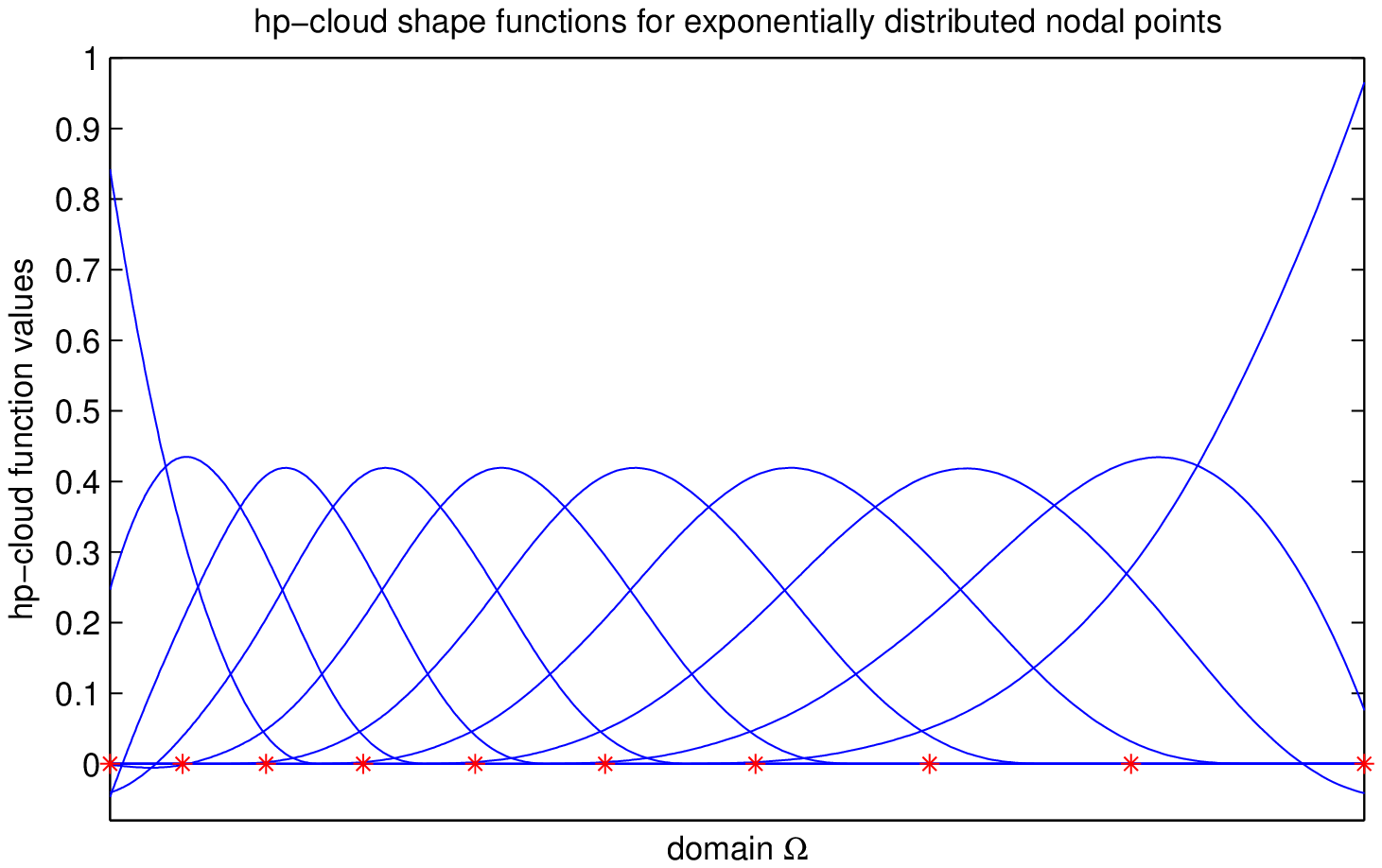}
\includegraphics[width=7.8cm]{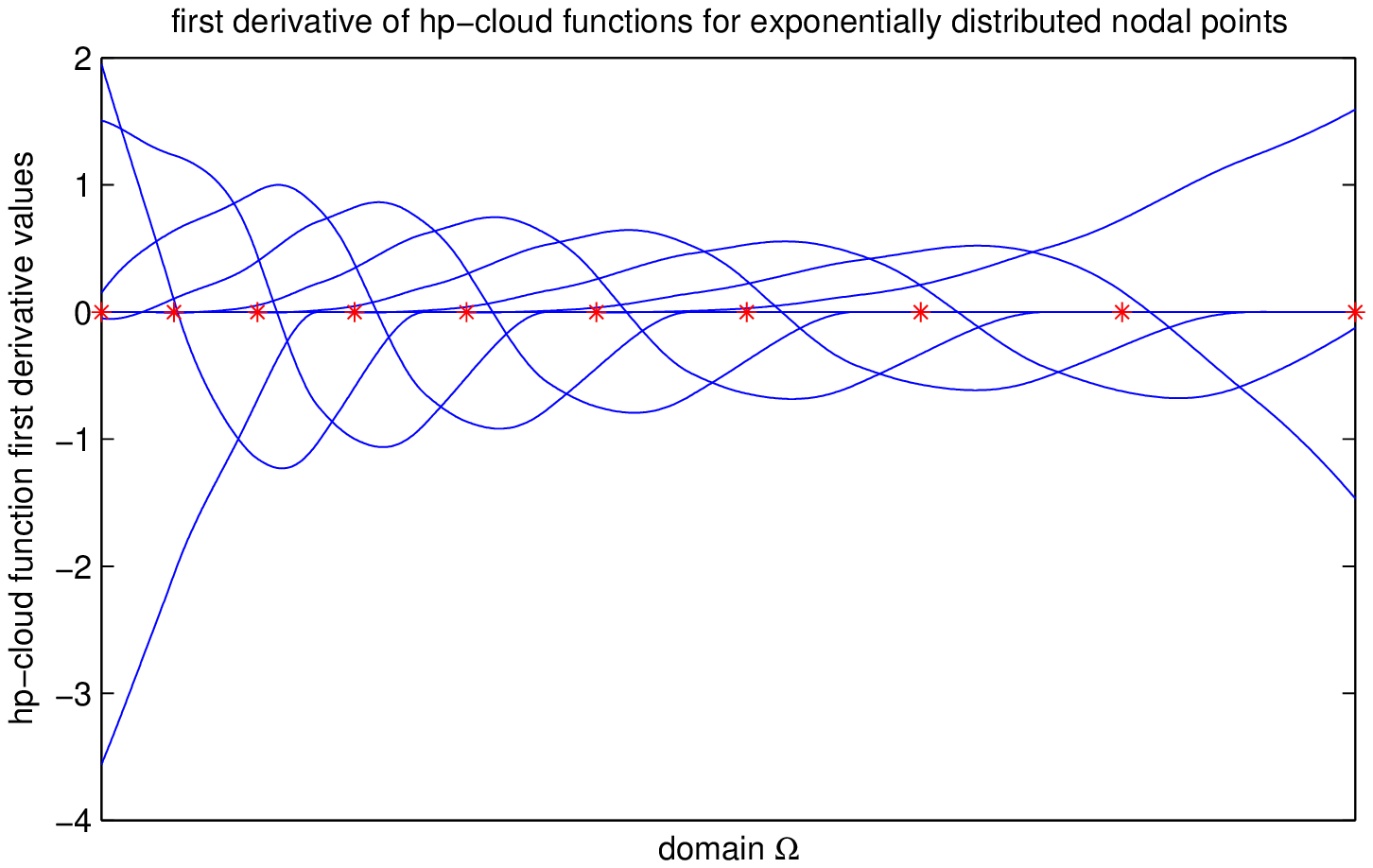}
\caption{PU $hp$-clouds (to the left) and their PN first derivatives (to the right). Quartic spline is used as a weight function.}
\end{figure}

Since the Kronecker delta property being not a character of $\psi_i$ ($\psi_i(x_j)\neq \delta_{ij}$), then at each node there are at least two nonzero shape functions. Thus, to have the value of the approximated function at a node, all nonzero shape functions values at that node should be added. The missing of the Kronecker delta property causes a problem in imposing EBCs, and thus other techniques are used to solve this difficulty, see below.

The intrinsic enrichment $P(x)$ has an important effect in the definition of the $hp$-cloud functions. All known fundamental characters, such as discontinuities and singularities, about the sought solution can be loaded on the intrinsic functions. Consequently, more time is saved; it is not needed, in general, to assume very large number of nodal points to capture a desired behavior of the approximated function while most of the solution features are inserted in the approximation itself. On the other hand, stability is enhanced particularly when there are some crucial characters that can not be captured by usual numerical methods, for example solving equations with rough coefficients that appear, e.g., in composites and materials with micro-structure, problems with high oscillatory solutions, or eigenvalue problems that admit spurious solutions in the computation of the discrete spectrum.\\

\textbf{Imposition of essential boundary conditions (EBCs)}\\

The radial Dirac eigenvalue problem assumes homogeneous Dirichlet boundary condition, while it is known that the $hp$-cloud approximation (MMs in general) can not treat this condition naturally, this is because the lack of the Kronecker delta property of the shape function. This is in contrast with most mesh-based methods, where the basis functions admit this property, and thus applying EBCs is straightforward (as in FEM) by omitting the first and the last basis functions.

In MMs in general, the widely applied techniques for imposing EBCs are Lagrangian multipliers, penalty condition, and coupling with finite element shape functions. Lagrangian multiplier is a very common and accurate approach for the imposition of EBCs. The disadvantage of this technique, see e.g. \cite{FRI2,ZHAN}, is that the resulted discrete equations for a self-adjoint operator are not positive definite (contains zero at the main diagonal) nor banded. Also the structure of the system becomes awkward, i.e., instead of having $\mathcal{M}$ as a resulting matrix of discretization of the Galerkin formulation, the system
$\left(
\begin{array}{cc}
\mathcal{M} & lm \\
lm & 0
\end{array}
\right)
$
is obtained, where $lm$ is the EBC-enforcement vector. EBCs can also be imposed by penalty condition \cite{FRI2,NOG}, the problem of applying this technique is the negative effect on the condition number of the resulting discrete equations.

The most powerful and safe method to enforce EBCs is coupling MMs with the FEM, applied for the first time in \cite{KRO}. With this approach, the meshfree shape functions of the nodes along boundaries are replaced by finite element basis functions. In one dimensional case, the $hp$-cloud functions at the first two and the last two nodes are replaced by finite element functions, and thus EBCs are, as in the FEM, simply imposed through eliminating the first and last added finite element functions, see Figure 2.
\begin{figure}[h]
\centering
\includegraphics[width=7.8cm]{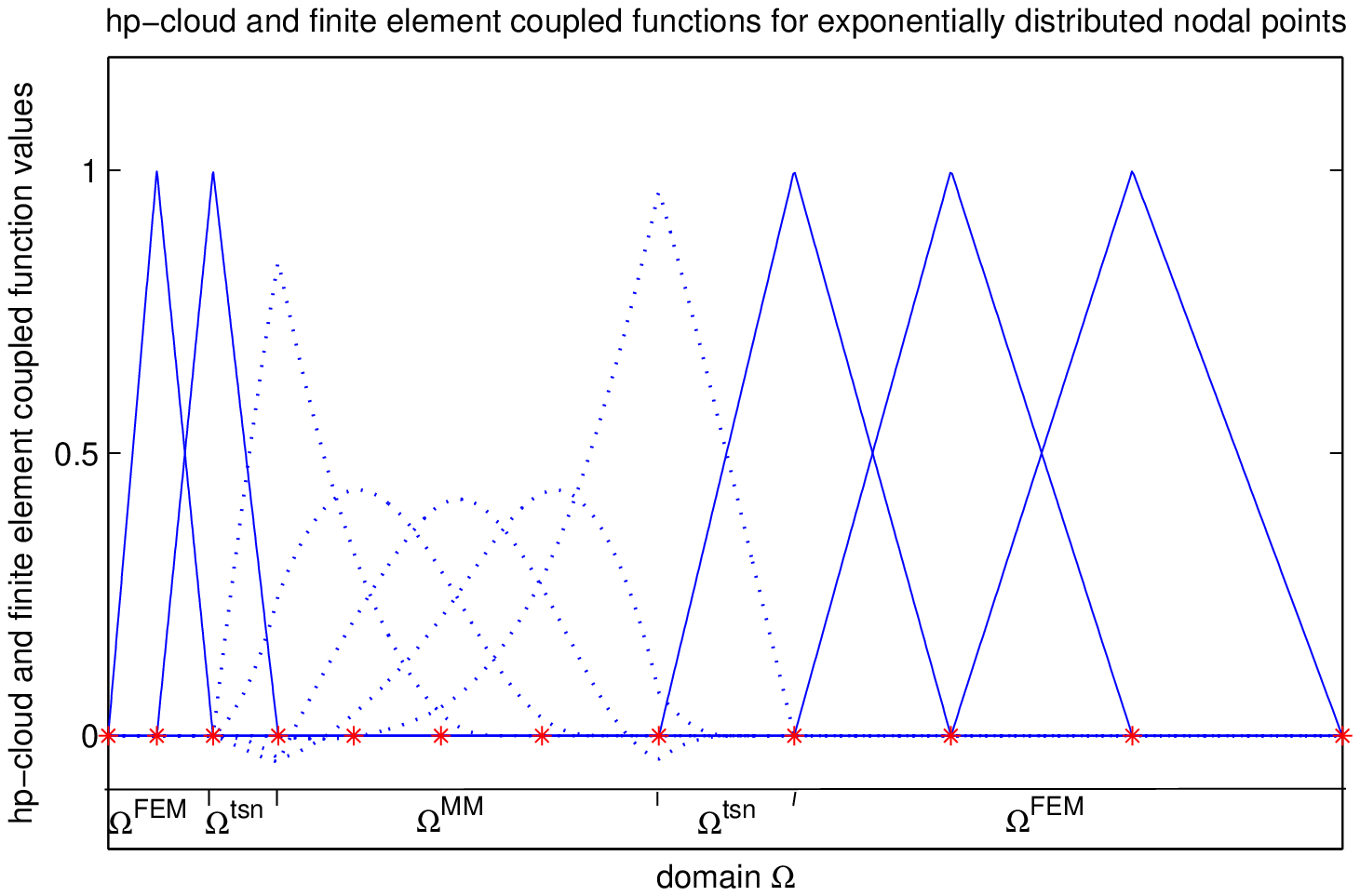}
\includegraphics[width=7.8cm]{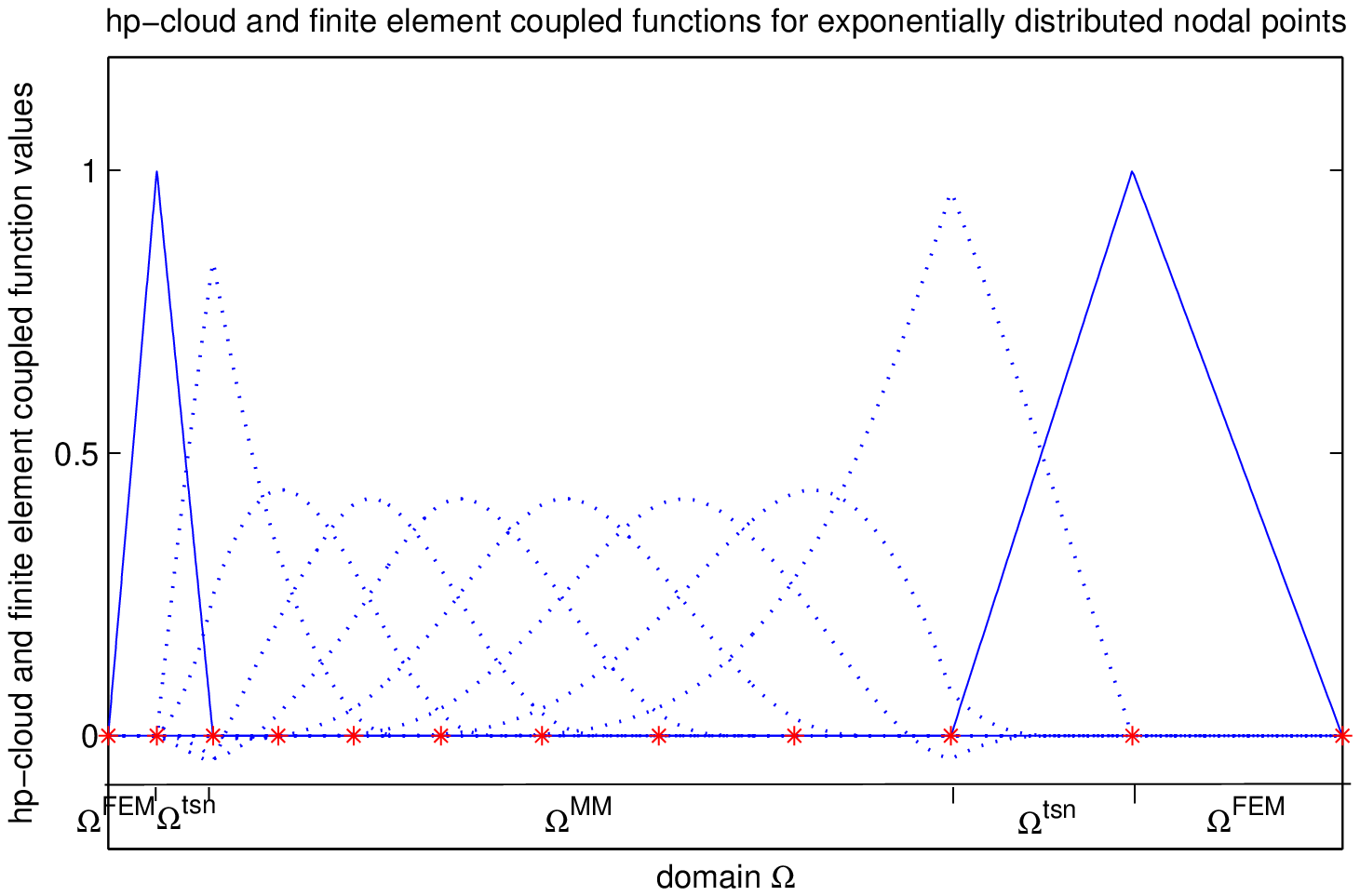}
\caption{Coupled $hp$-cloud and finite element functions: general coupling (to the left), and coupling for the purpose of imposing EBCs (to the right) (two finite element functions are sufficient). Linear (hat) functions are used as finite element functions, and quartic spline as a weight function in the $hp$-clouds.}
\end{figure}

Two efficient approaches of coupling MMs with the FEM are coupling with Ramp functions \cite{BEL}, and coupling with reproducing conditions \cite{HUE}. With the former approach, the derivative of the resulting coupled approximation function at the boundary of the interface region, $\Omega^{\textsf{tsn}}$ in Figure 2, is discontinuous and the consistency is of first order. To ensure the continuity of the first derivative of the coupled function and to obtain consistency of any order, we consider the latter approach. Using MLS approximation method as before, the approximation resulting from coupling $hp$-cloud and finite element functions with the reproducing conditions is given as (see e.g. \cite{FRI2})
\begin{eqnarray*}
\begin{array}{ll}
\Psi_h(x)\!\!\!\!\!&=\!\!\!\!\!\displaystyle \sum_{x_i\in\Omega^{\textsf{MM}}}\!\!\!\!\! \psi_i(x) \Psi_i +\!\!\!\!\! \displaystyle \sum_{x_i\in\Omega^{\textsf{FEM}}}\!\!\!\!\!\! \mathcal{G}_i(x)\Psi_i\\
 \!\!\!\!\!&=\!\!\!\!\!\displaystyle\sum_{x_i\in\Omega^{\textsf{MM}}}\!\!\!\!\! \Big(P^t(x)-\!\!\!\!\!\sum_{x_i\in\Omega^{\textsf{FEM}}}\!\!\!\!\!\! \mathcal{G}_i(x)P^t(x_i)\Big) M^{-1}(x) \varphi_i(\frac{x-x_i}{\rho_i})P(x_i)\Psi_i +\!\!\!\!\! \displaystyle \sum_{x_i\in\Omega^{\textsf{FEM}}}\!\!\!\!\!\! \mathcal{G}_i(x)\Psi_i,
\end{array}
\end{eqnarray*}
where $\mathcal{G}_i$ are the finite element shape functions, and $M$ is as defined before. From Figure 2, it can be seen that finite element functions are only complete in $\Omega^{\textsf{FEM}}$, and that in $\Omega^{\textsf{MM}}$ only $hp$-clouds are present. In the transition interface region, $\Omega^{\textsf{tsn}}$, the existence of incomplete finite element functions modifies the existed $hp$-clouds there, and thus coupled $hp$-cloud and finite element functions are obtained.

\section{The scheme and the stability parameter}

Since the radial Dirac eigenvalue problem is convection dominated, the $hp$-cloud approximation of it will be unstable. As most of applications of numerical methods, certain modifications are used to stabilize solutions \cite{ALMA,ALME,BRO81,BRO82,DES,IDE}. Therefore, instead of considering the $hp$-cloud approximation for the radial Dirac eigenvalue problem, we will apply the $hp$-CPG method to create diffusion terms to stabilize the approximation. The $hp$-CPG method is a consistent method in the sense that the solution of the original problem is also a solution to the weak form. The size of the added diffusivity is controlled by a stability parameter. To set the scheme, consider the radial Dirac eigenvalue problem $H_\kappa \Phi=\lambda\Phi$, the usual $hp$-cloud method is formulated by multiplying the equation by a test function $\Psi$ and integrating over the domain $\Omega$
\begin{equation}\label{59}
\int_{\Omega}\Psi^t H_\kappa \Phi dx=\lambda\int_{\Omega}\Psi^t \Phi dx\, .
\end{equation}
To discretize (\ref{59}) let $\Psi$ be defined as $(\psi_i,0)^t$ and $(0,\psi_i)^t$, $i=1, 2, \ldots, n$, where $\psi_i$ is the $hp$-cloud basis function and
\begin{equation}\label{60}
\Phi(x) = \left(
\begin{array}{c}
F(x) \\
G(x)
\end{array}
\right)
= \left(
\begin{array}{c}
\sum_{j=1}^{n}f_j\psi_j(x) \\
\sum_{j=1}^{n}g_j\psi_j(x)
\end{array}
\right)\, .
\end{equation}
The elements $f_j$ and $g_j$ are the nodal values of $F$ and $G$ respectively. This yields

\begin{equation}\label{21}
\sum_{j=1}^n\langle w^{+}(x)\psi_j(x)\, ,\, \psi_i(x)\rangle f_j\!+\!\sum_{j=1}^n\langle -c\psi_j'(x)+\frac{c\kappa}{x}\psi_j(x)\, , \, \psi_i(x)\rangle g_j=\lambda\sum_{j=1}^n\langle\psi_j(x)\, ,\, \psi_i(x)\rangle f_j
\end{equation}
and
\begin{equation}\label{22}
\sum_{j=1}^n\langle c\psi_j'(x)+\frac{c\kappa}{x}\psi_j(x)\, , \, \psi_i(x)\rangle f_j \!+\!\sum_{j=1}^n\langle w^{-}(x)\psi_j(x)\, ,\, \psi_i(x)\rangle g_j=\lambda\sum_{j=1}^n\langle \psi_j(x)\, ,\, \psi_i(x)\rangle g_j\, ,
\end{equation}
the bracket $\langle\cdot\,,\,\cdot\rangle$ is the usual $L^2(\Omega)$ scalar product. After simplifying, equations (\ref{21}) and (\ref{22}) lead to the symmetric generalized eigenvalue problem
\begin{equation}\label{23}
AX=\lambda BX\, .
\end{equation}
The block matrices $A$ and $B$ are defined by
\begin{equation}\label{24}
A=\left(\begin{array}{c|c}
mc^2M_{000}+M_{000}^V & -cM_{010}+c\kappa M_{001} \\
\hline
cM_{010}+c\kappa M_{001} & -mc^2M_{000}+M_{000}^V
\end{array}\right), \;
\text{and}\;
B=\left(\begin{array}{c|c}
M_{000}& 0 \\
\hline
0 & M_{000}
\end{array}\right)\, ,
\end{equation}
where $M_{rst}^q$ are $n\times n$ matrices given as
\begin{equation}\label{26}
(M_{rst}^q)_{ij}=\int_{\Omega}\psi_j^{(s)}\,\psi_i^{(r)}\,x^{-t}\,q(x)\,dx\; ,\;\;\;\;\Big(\psi^{(r)}(x)=\frac{d^r}{dx^r}\psi (x)\Big)\, .
\end{equation}
The vector $X$ is the unknowns defined as $(f\, ,\, g)^t$, where $f=(f_1,f_2,\ldots,f_n)$ and $g=(g_1,g_2,\ldots,g_n)$.

To formulate the $hp$-CPG method, the test function $\Psi$ is modified to include the first derivative of the basis function in order to introduce the required diffusivity. This leads to assume $\Psi$ as $(\psi,\tau \psi')^t$ and $(\tau \psi', \psi)^t$ in (\ref{59}), where $\tau$ is the stability parameter that controls the size of the diffusion terms, $\psi=\psi_i$, and the functions $F$ and $G$ are given by (\ref{60}), thus we get
\begin{equation}\label{61}
\langle w^{+}F\, ,\, \psi\rangle+\langle-cG'+\frac{c\kappa}{x}G\, ,\, \psi\rangle+\langle Re^2(F,G)\, ,\, \tau \psi'\rangle=\lambda\langle F\, ,\, \psi\rangle
\end{equation}
and
\begin{equation}\label{62}
\langle cF'+\frac{c\kappa}{x}F\, ,\, \psi\rangle+\langle w^{-}G\, ,\, \psi\rangle+\langle Re^1(F,G)\, ,\, \tau \psi'\rangle=\lambda\langle G\, ,\, \psi\rangle\, .
\end{equation}
The residuals $Re^{1}\big(F,G\big)(x)$ and $Re^{2}\big(F,G\big)(x)$ are defined as
\begin{equation}\label{57}
Re^{1}\big(F,G\big)(x)=\big(W^{+}F-cG'+\frac{c\kappa}{x}G\big)(x)\, ,
\end{equation}
\begin{equation}\label{58}
Re^{2} \big(F,G\big)(x)=\big(W^{-}G+cF'+\frac{c\kappa}{x}F\big)(x)\, ,
\end{equation}
where $W^{\pm}(x)=w^{\pm}(x)-\lambda$. This results in the usual $hp$-cloud approximation with addition to perturbations sized by $\tau$ as follows
\begin{equation}\label{64}
\mathbf{A}X=\lambda\mathbf{B}X\,.
\end{equation}
The perturbed block matrices, $\mathbf{A}$ and $\mathbf{B}$, are respectively in the forms $\mathbf{A}=A+\tau \mathcal{A}$ and $\mathbf{B}=B+\tau \mathcal{B}$, where $A$ and $B$ are given by (\ref{24}),
\begin{equation}\label{63}
\mathcal{A}=\left(\begin{array}{c|c}
c M_{110}+c\kappa M_{101} & -mc^2 M_{100}+ M_{100}^V\\
\hline
mc^2 M_{100}+ M_{100}^V & -c M_{110}+c\kappa M_{101}
\end{array}\right),\;
\text{and}\;
\mathcal{B}=\left(\begin{array}{c|c}
0 &  M_{100} \\
\hline
 M_{100} & 0
\end{array}\right)\, .
\end{equation}

The system (\ref{64}) is not symmetric, thus complex eigenvalues may appear if the size of $\tau$ is relatively large. In the FEM, an explicit representation for $\tau$ is obtained \cite{ALMA}, where the basis functions have the Kronecker delta property, hence the basis functions have regular distribution along the domain and only the adjacent basis functions intersect in one and only one subinterval. Thus the resulted system consists of tridiagonal matrices, this makes the derivation of $\tau$ easier and an explicit representation is feasible. In MMs in general, a basis function is represented by cloud over a nodal point, with domain of influence, $\rho$, that may cover many other nodal points. So the resulting matrices can be filled with many nonzero elements, hence the number of non-vanishing diagonals in these matrices is arbitrary (greater than 3) and depending on the size of $\rho$. Therefore, we can not write an explicit representation for $\tau$ that depends only on a given mesh. Instead, $\tau$ will be mainly represented by some of the computed matrices obtained from the usual $hp$-cloud method.

The derivation of $\tau$ assumes the limit Dirac operator in the vicinity of $x$ at infinity. This presumable simplification is inevitable and justifiable; the derivation leads to an approximation of the limit point eigenvalue which depends on $\tau$, where, in \cite{GRI}, the theoretical limit is proved to be $mc^2$, hence we can minimize the error between the theoretical and the approximated limits to get $\tau$. By considering the limit operator at infinity, we consider the troublesome part (that includes the convection terms) of the operator which is mostly needed to be stabilized. Besides that, one is obliged to assume that the stability parameter should be applicable at all radial positions $x\in\Omega$, particularly the large values of $x$.
\begin{Theo}
Let $M_{000}$ and $M_{100}$ be the $n\times n$ computed matrices given by (\ref{26}), and let $\sigma_{ji}$ and $\eta_{ji}$ be the corresponding entries respectively. Define $\vartheta$ as
\begin{equation}\label{1000}
\vartheta_{ji}=\left\{
\begin{array}{ll}
\!\!\!\displaystyle-\!\!\!\sum_{k=i+1}^{j}h_k\,,&i<j,\\
\quad0\,,&i=j,\\
\!\!\!\displaystyle\sum_{k=j+1}^{i}h_k\,,&i>j,
\end{array}
\right.
\end{equation}
where $h_k$ is the displacement between the nodes $x_k$ and $x_{k-1}$. Then the stability parameter, $\tau_j$, for an arbitrary $j^{th}$ row of the matrices in $\mathcal{A}$ and $\mathcal{B}$ is given by
\begin{equation}\label{100}
\tau_j=\Big|\displaystyle\sum_{i=1}^{n}\sigma_{ji}\vartheta_{ji}\Big/\displaystyle
\sum_{i=1}^{n}\eta_{ji}\vartheta_{ji}\Big|.
\end{equation}
\end{Theo}
\hspace{-4mm}\underline{\emph{Proof}}. Consider the limit operator of the radial Dirac eigenvalue problem in the vicinity of $x$ at infinity
\begin{equation}\label{70}
\left(
\begin{array}{cc}
mc^2 & -cD_x \\
cD_x & -mc^2
\end{array}
\right)
 \left(
\begin{array}{c}
F(x) \\
G(x)
\end{array}
\right)
=
 \lambda \left(
\begin{array}{c}
F(x) \\
G(x)
\end{array}
\right)\, .
\end{equation}
The $hp$-CPG variational formulation of (\ref{70}) (which is equivalent to assume a limit passage as $x\to\infty$ of the equations (\ref{61}) and (\ref{62})) provides
\begin{eqnarray}\label{66}
(mc^2-\lambda)M_{000}f+\tau cM_{110}f-(\tau mc^2-c+\tau\lambda) M_{100}g=0
\end{eqnarray}
and
\begin{eqnarray}\label{67}
(\tau mc^2-c-\tau\lambda)M_{100}f-\tau cM_{110}g-(mc^2+\lambda) M_{000}g=0\,,
\end{eqnarray}
where, as defined before, $f=(f_1,f_2,\ldots,f_n)$ and $g=(g_1,g_2,\ldots,g_n)$. Let $\sigma_{k}$, $\eta_{k}$, and $\varrho_{k}$, for $k=1,2,\ldots, n$, be the corresponding $j^{th}$ row entries of $M_{000}$, $M_{100}$, and $M_{110}$ respectively. To obtain $\tau_j$, we consider the $j^{th}$ rows in (\ref{66}) and (\ref{67}), this together with the Lemma 1 below gives
\begin{align}\label{68}
\Big(mc^2\!-\!\lambda\Big)\Big(\displaystyle \sum_{k=1}^n\sigma_kf_j \!+\!\displaystyle
\sum_{k=1}^n\sigma_k
\big(mc\vartheta_k\!+\!(\vartheta_k/c)\lambda\big)g_j\Big)\!+\!\tau c\Big(\displaystyle\sum_{k=1}^n\displaystyle\varrho_kf_j\!
+\!\displaystyle\sum_{k=1}^n\displaystyle\varrho_k
\big(mc\vartheta_k+\\\nonumber
+(\vartheta_k/c)\lambda\big)g_j\Big)-\Big(\tau mc^2-c+\tau\lambda\Big) \Big(\displaystyle\sum_{k=1}^n\eta_kg_j+\displaystyle\sum_{k=1}^n\eta_k
\big(mc\vartheta_k-(\vartheta_k/c)\lambda\big)f_j\Big)=0
\end{align}
and
\begin{align}\label{69}
\Big(\tau mc^2-c-\tau\lambda\Big)\Big(\displaystyle \sum_{k=1}^n\eta_kf_j+\displaystyle
\sum_{k=1}^n\eta_k
\big(mc\vartheta_k+(\vartheta_k/c)\lambda\big)g_j\Big)-
\tau c \Big(\displaystyle\sum_{k=1}^n\varrho_kg_j+\\\nonumber
+\displaystyle\sum_{k=1}^n\varrho_k
\big(mc\vartheta_k
-(\vartheta_k/c)\lambda\big)f_j\Big)-(mc^2+\lambda)\Big(\displaystyle
\sum_{k=1}^n\sigma_kg_j+\displaystyle\sum_{k=1}^n\sigma_k
\big(mc\vartheta_k-(\vartheta_k/c)\lambda\big)f_j\Big) =0\,.
\end{align}
\begin{Lem}
Let $f_i$ and $g_i$ be respectively the $i^{th}$ nodal values of $F$ and $G$ of the limit equation (\ref{70}). Freeze $j$, and let $\vartheta_i$ be given by (\ref{1000}) for the given $j$. Then for $i=1,2,\ldots, n$
\begin{eqnarray*}
 && f_i\approx f_j+\Big(mc\vartheta_i+(\vartheta_i/c)\lambda\Big)g_j\, . \\
 && g_i\approx g_j+\Big(mc\vartheta_i-(\vartheta_i/c)\lambda\Big)f_j\, .
\end{eqnarray*}
\end{Lem}
\hspace{-4mm}\underline{\emph{Proof}}. Consider the limit equation (\ref{70}) which can be written as
\begin{eqnarray}\label{111}
mc^2F(x)-c G'(x)=\lambda F(x)\;\text{ and }\;
c F'(x)-mc^2G(x)=\lambda G(x)\,.
\end{eqnarray}
If $i=j$, then the result is obvious. So let $i\neq j$, we treat the case $i<j$, where the proof for $i>j$ goes through mutatis mutandis by using forward difference approximations for derivatives. Assume $i<j$, also we prove the first assertion of the lemma, the proof of the second assertion is similar. Consider the second part of (\ref{111}) for $x_j$
\begin{equation}\label{112}
c F'(x_j)-mc^2G(x_j)=\lambda G(x_j)\,.
\end{equation}
Using backward difference approximations for derivatives we can write
\begin{equation}\label{113}
F'|_{x_j}\approx \frac{F(x_j)-F(x_{i})}{\displaystyle-\!\!\sum_{k=i+1}^{j}h_k}= \frac{f_j-f_i}{\displaystyle-
\!\!\sum_{k=i+1}^{j}h_k}.
\end{equation}
Substituting (\ref{113}) in (\ref{112}) completes the proof.\hfill{$\blacksquare$}\\

We continue the proof of Theorem 1, consider the dominant parts with respect to $c$, so let $c\to\infty$ in (\ref{68}) and (\ref{69}) and simplify to get
\begin{eqnarray}\label{71}
\Big[\displaystyle\sum_{k=1}^n\Big(\big(-\sigma_k -\eta_k\vartheta_k\big)\lambda+\big(c\varrho_k-m^2c^3\eta_k\vartheta_k\big)\tau_j
+\big(mc^2\sigma_k+mc^2\eta_k\vartheta_k\big)\Big)\Big]f_j+\\\nonumber
+\Big[\displaystyle\sum_{k=1}^n\Big(\big(\tau_j\varrho_k \vartheta_k-\tau_j\eta_k\big)\lambda
+\big(mc^2\varrho_k\vartheta_k-mc^2\eta_k\big)\tau_j+\big(m^2c^3\sigma_k\vartheta_k
+c\eta_k\big)\Big)\Big]g_j=0
\end{eqnarray}
and
\begin{eqnarray}\label{72}
\Big[\displaystyle\sum_{k=1}^n\Big(\big(-\tau_j\eta_k+\tau_j \varrho_k\vartheta_k\big)\lambda
+\big(mc^2\eta_k-mc^2\varrho_k\vartheta_k\big)\tau_j+\big(-c\eta_k-m^2c^3\sigma_k
\vartheta_k\big)\Big)\Big]f_j+\\\nonumber
+\Big[\displaystyle\sum_{k=1}^n\Big(\big(-\eta_k\vartheta_k-\sigma_k\big)\lambda
+\big(m^2c^3\eta_k\vartheta_k-c\varrho_k\big)\tau_j
+\big(-mc^2\eta_k\vartheta_k-mc^2\sigma_k\big)\Big)\Big]g_j=0.
\end{eqnarray}
To make the derivation simpler, the following notations are introduced
$$
\begin{array}{ll}
a=\sum_{k=1}^{n}a_k=\sum_{k=1}^{n}(-\sigma_k-\eta_k\vartheta_k), & b=cb_1-m^2c^3b_2=\sum_{k=1}^{n}(c\varrho_k-m^2c^3\eta_k\vartheta_k),\\
d=mc^2d_1=\sum_{k=1}^{n}mc^2(\sigma_k+\eta_k\vartheta_k),& e=\sum_{k=1}^{n}e_k=\sum_{k=1}^{n}(\varrho_k\vartheta_k-\eta_k),\\
q=mc^2q_1=\sum_{k=1}^{n}mc^2(\varrho_k\vartheta_k-\eta_k), & \omega=m^2c^3\omega_1+c\omega_2= \sum_{k=1}^{n}(m^2c^3\sigma_k\vartheta_k+c\eta_k).
\end{array}
$$
By these notations, equations (\ref{71}) and (\ref{72}) can be written as
\begin{equation}\label{73}
\left(
\begin{array}{cc}
a\lambda+b\tau_j+d & e\tau_j\lambda+q\tau_j+\omega \\
e\tau_j\lambda-q\tau_j-\omega & a\lambda-b\tau_j-d
\end{array}
\right)
 \left(
\begin{array}{c}
f_j \\
g_j
\end{array}
\right)
=
  \left(
\begin{array}{c}
0 \\
0
\end{array}
\right)\, .
\end{equation}
Since $f_j$ and $g_j$ are not identically zero for all $j$, then we expect
\begin{equation}\label{74}
det\left(
\begin{array}{cc}
a\lambda+b\tau_j+d & e\tau_j\lambda+q\tau_j+\omega \\
e\tau_j\lambda-q\tau_j-\omega & a\lambda-b\tau_j-d
\end{array}
\right)=0,
\end{equation}
where $det(\cdot)$ is the determinant of matrix. After simplifying, equation (\ref{74}) leads to
\begin{equation}\label{75}
\lambda_{\pm}(\tau_j)=\pm\sqrt{\frac{(b\tau_j+d)^2-(q\tau_j+\omega)^2}{a^2-e^2\tau_j^2}}.
\end{equation}
By \cite{GRI}, the only accumulation point for the eigenvalue for the radial Coulomb-Dirac operator in the vicinity of $x$ at infinity is $mc^2$. So, we like to have
\begin{eqnarray*}
\begin{array}{lll}
                      \vspace{1mm}
                     & |\lambda_{+}-mc^2|=0&\\
                     \vspace{1mm}
 \Longleftrightarrow & m^2c^4(a^2- e^2\tau_j^2)&=(b\tau_j+d)^2-(q\tau_j+ \omega)^2\\
                     &                      &=(cb_1\tau_j-m^2c^3b_2\tau_j +mc^2d_1)^2- (mc^2q_1\tau_j+m^2c^3\omega_1 +c\omega_2)^2.
\end{array}
\end{eqnarray*}
Letting $m=1$, dividing both sides by $c^6$, and taking the limit as $c\to\infty$, we get
\begin{equation}\label{76}
b_2^2\tau_j^2-\omega_1^2=0.
\end{equation}
Substituting back the values of $b_2$ and $\omega_1$, the desired consequence is obtained for the fixed $j$ as
\begin{equation}\label{77}
\tau_j=\Big|\displaystyle\sum_{k=1}^n\sigma_k\vartheta_k\Big/\displaystyle
\sum_{k=1}^n\eta_k\vartheta_k\Big|.
\end{equation}
The above result can be generalized for arbitrary $j$ as
\begin{equation}\label{78}
\tau_j=\Big|\displaystyle\sum_{i=1}^{n}\sigma_{ji}\vartheta_{ji}\Big/\displaystyle
\sum_{i=1}^{n}\eta_{ji}\vartheta_{ji}\Big|,
\end{equation}
and this ends the proof.\hfill{$\blacksquare$}\\

The $hp$-cloud functions depend strongly on the dilation parameter $\rho_j$. As $\rho_j$ gets smaller, i.e., $\rho_j\rightarrow max\{h_j,h_{j+1}\}$ ($=h_{j+1}$ for exponentially distributed nodal points), as the shape functions of MMs in general become closer to the standard finite element functions, see Figure 3. In this case the FEM stability parameter might be applicable for MMs \cite{FRI1}
$$
\tau_j^{F\!E\!M}\rightarrow \tau_j^{M\!M\!s}\,,\quad \text{as} \quad \rho_j\rightarrow h_{j+1}.
$$
\begin{figure}[h]
\centering
\includegraphics[width=7.8cm]{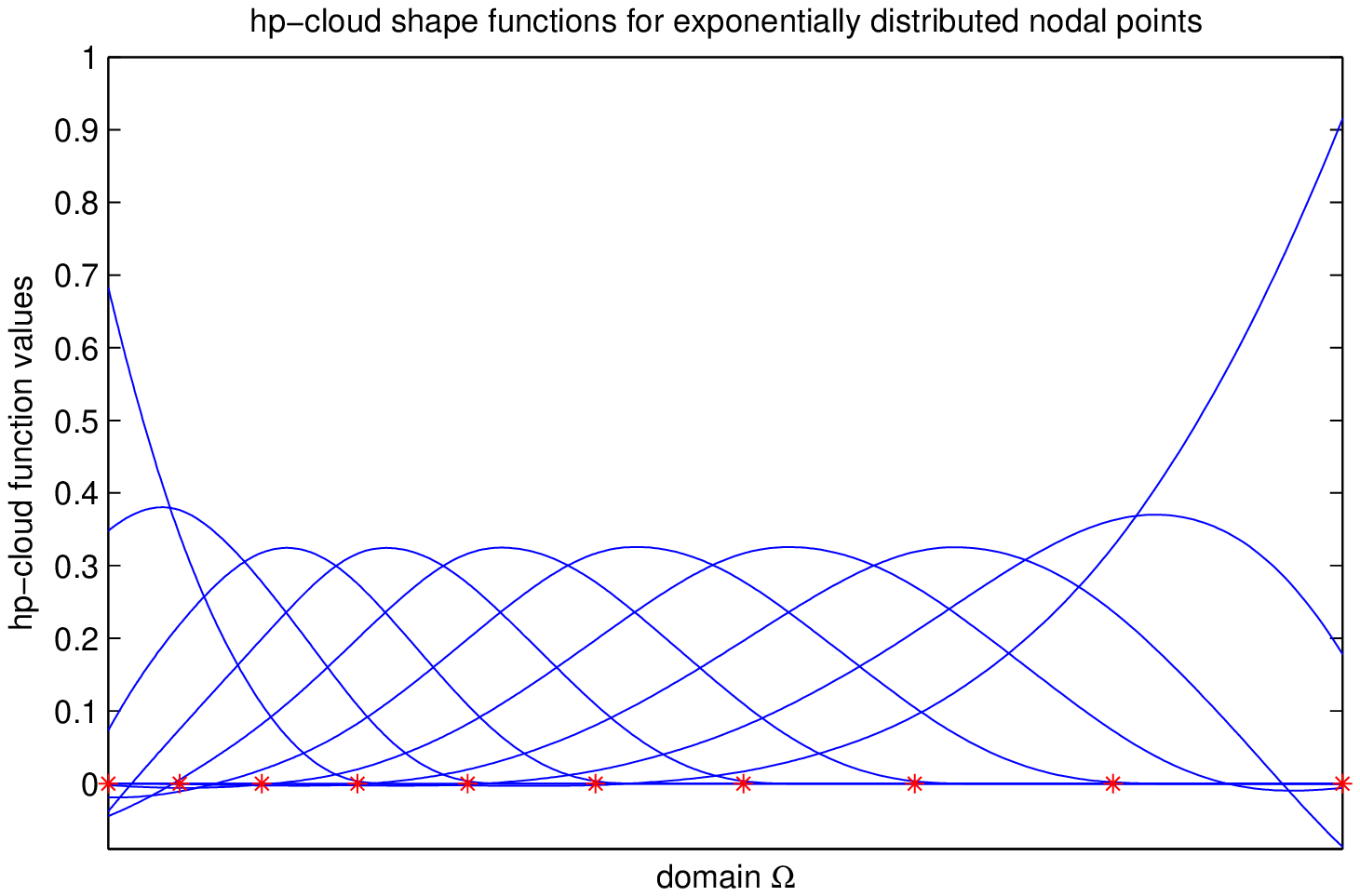}
\includegraphics[width=7.8cm]{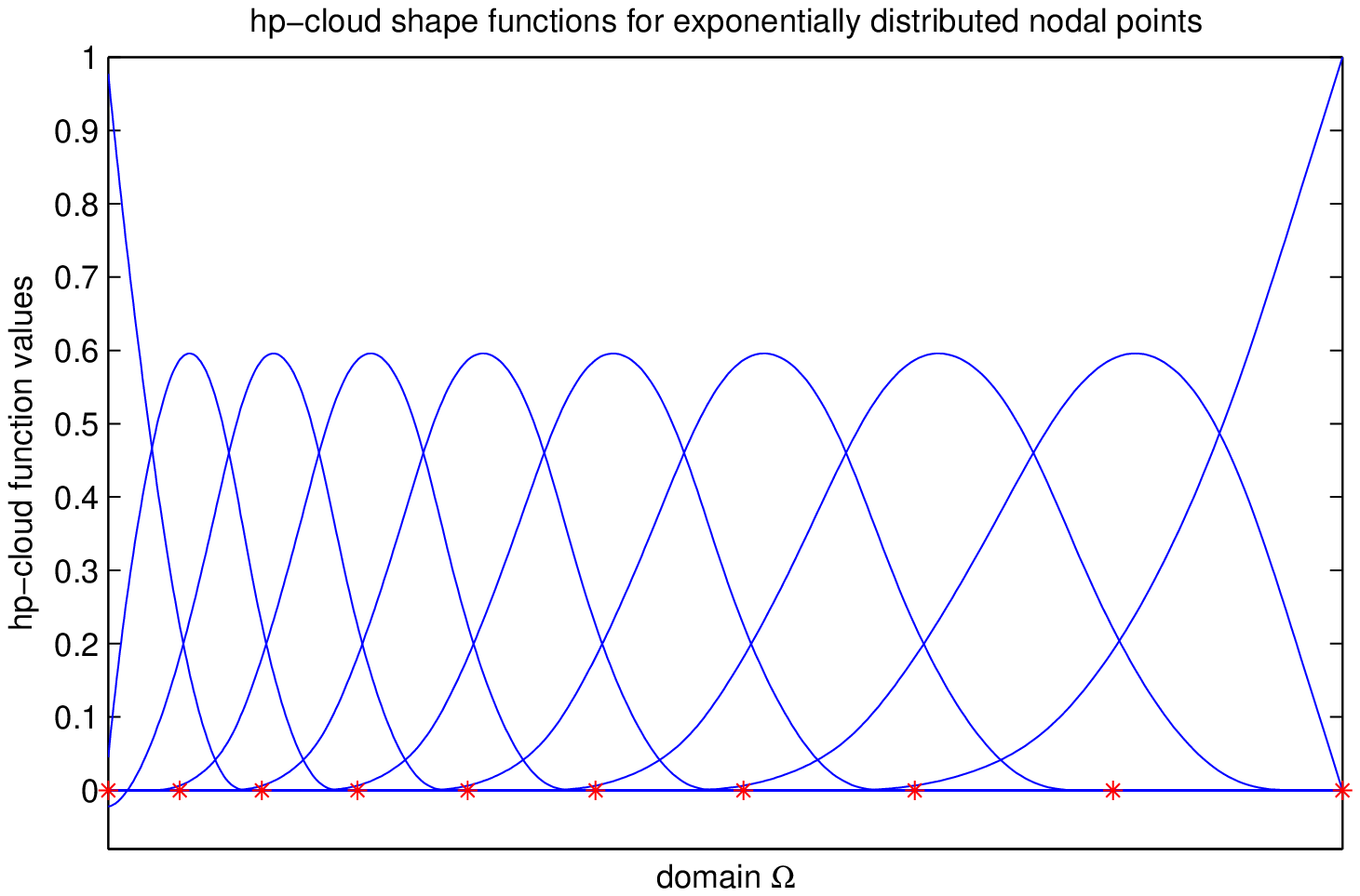}\\
\includegraphics[width=7.8cm]{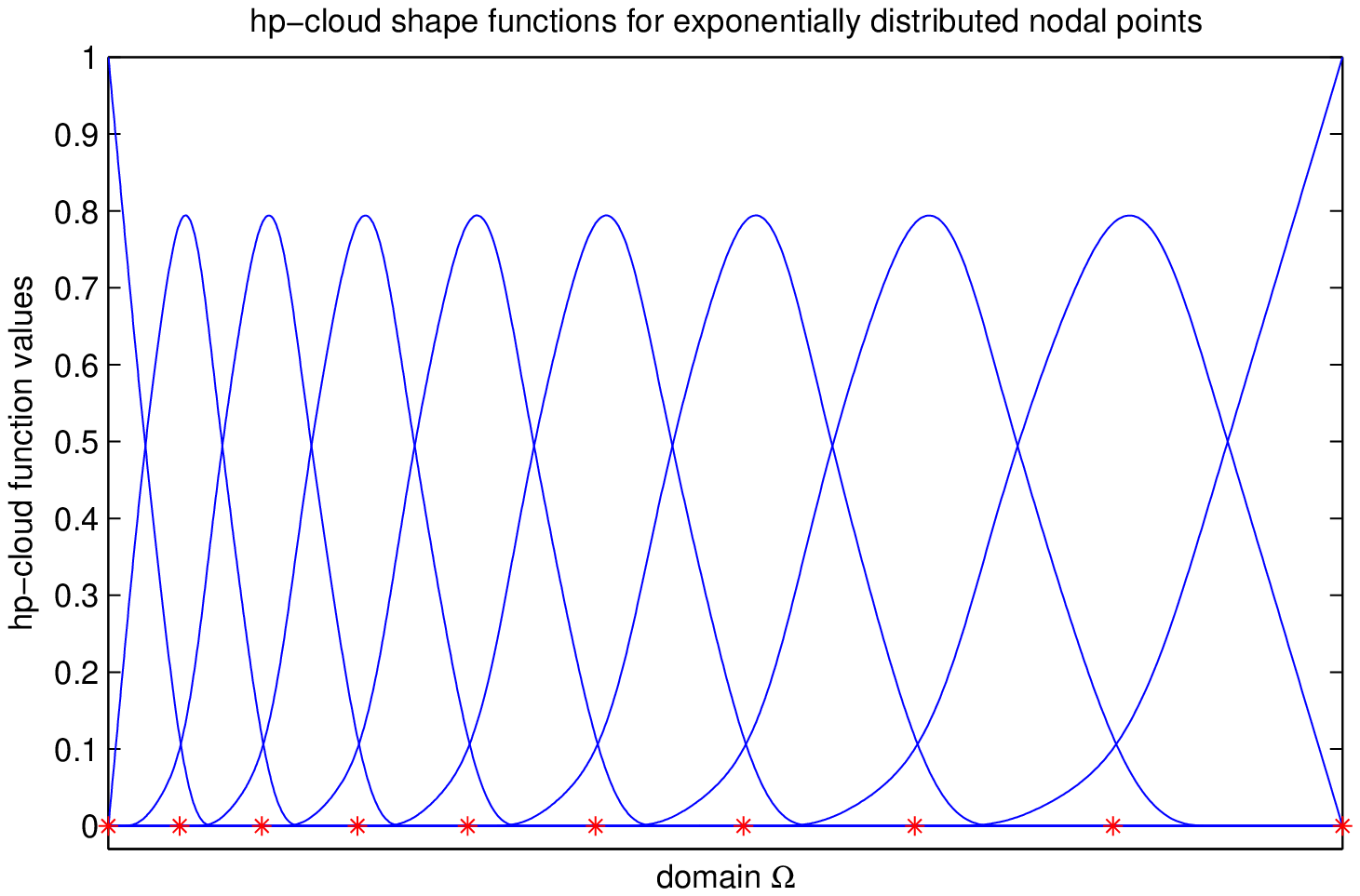}
\includegraphics[width=7.8cm]{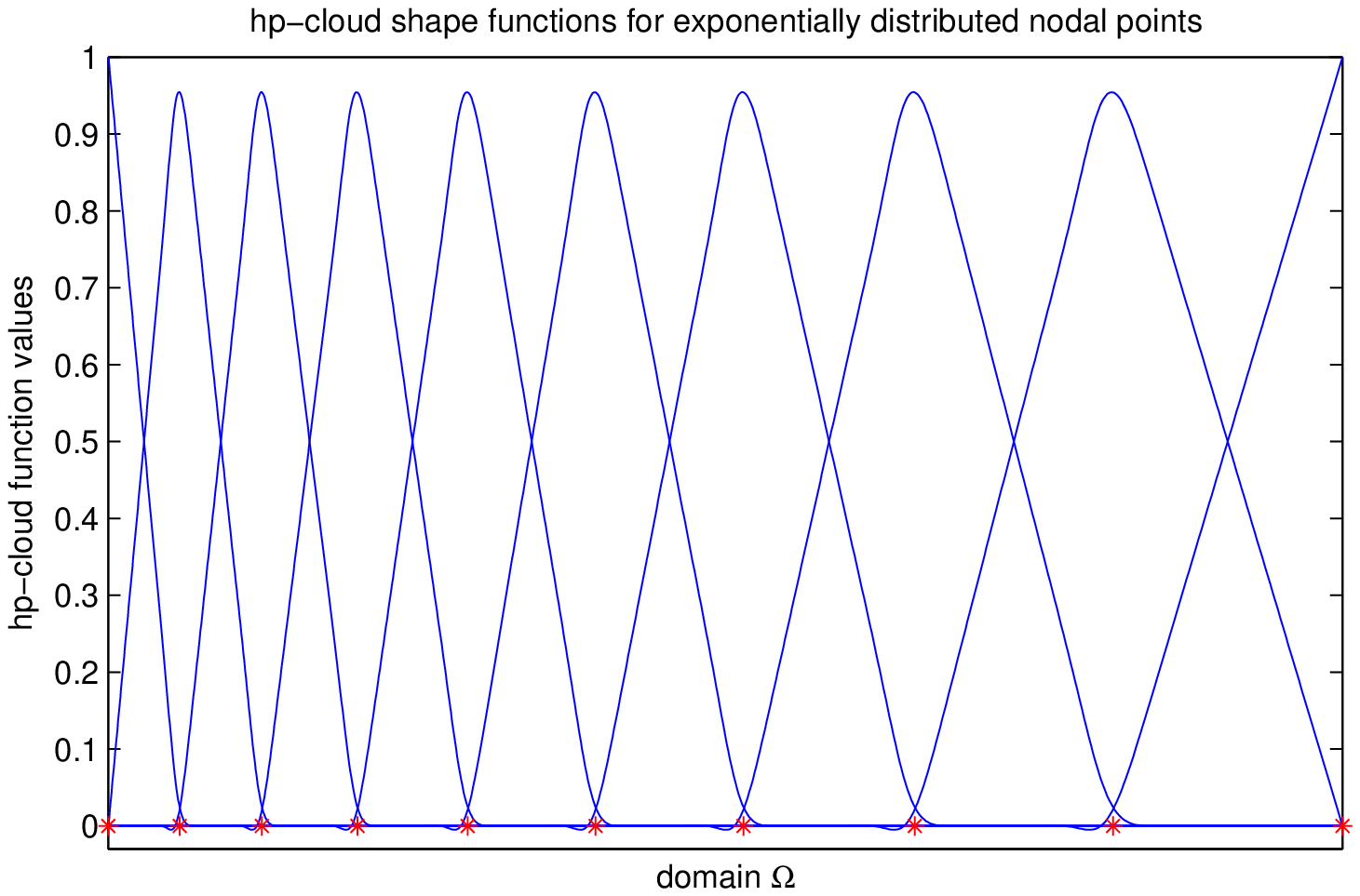}
\caption{PU $hp$-clouds with different dilation parameters: $\rho_j=4\cdot h_{j+1}$ (up to the left), $\rho_j=2\cdot h_{j+1}$ (up to the right), $\rho_j=1.5\cdot h_{j+1}$ (below to the left), and $\rho_j=1.2\cdot h_{j+1}$ (below to the right). Quartic spline is used as a weight function.}
\end{figure}
On the other hand, one should be careful about the invertibility of the matrix $M$, i.e., we can not approach $\rho_j=h_{j+1}$ which makes $M$ singular. In Lemma 2, we derive the stability parameter $\tau_j^{F\!E\!M}$ for the computation of the eigenvalues of the radial Dirac operator, $H_\kappa$, using the FEM with quartic spline. The proof of the lemma is simple and uses the same technique as of the theorem above, thus we directly utilize the result of this theorem with minor modifications. In Table 7, the result of applying $\tau_j^{F\!E\!M}$ for stabilizing the $hp$-cloud method with $\rho_j=1.1\cdot h_{j+1}$ is obtained, the approximation is good enough and the spuriosity seems to be eliminated. But a difficulty arises, that is, the end of the spectrum (the spectrum tail) behaves in a strange way, which may be regarded as spurious solutions.
\begin{Lem}
The FEM stability parameter for the computation of the eigenvalues of the radial Dirac operator using quartic splines as a basis has the form
\begin{equation}\label{79}
\tau_j^{F\!E\!M}=\frac{3}{17}h_{j+1}\frac{(h_{j+1}-h_j)}{(h_{j+1}+h_j)}\, .
\end{equation}
\end{Lem}
\hspace{-4mm}\underline{\emph{Proof}}. Consider the general formula derived in Theorem 1
\begin{equation}\label{80}
\tau_j=\Big|\displaystyle\sum_{i=1}^{n}\sigma_{ji}\vartheta_{ji}\Big/\displaystyle
\sum_{i=1}^{n}\eta_{ji}\vartheta_{ji}\Big|,
\end{equation}
where $\vartheta_{ji}$ is defined by (\ref{1000}), and $\sigma_{ji}$ and $\eta_{ji}$ are respectively the entries of the matrices $M_{000}$ and $M_{100}$. Note that in the FEM with quartic spline basis functions, $M_{000}$ and $M_{100}$ are tridiagonal matrices with $j^{th}$ row elements as in Table 2.
\begin{table}[ht]
\caption{The element integrals of the matrices $M_{000}$ and $M_{100}$.}
\centering
\begin{tabular}{@{} ||c|c|c|c|| @{} }
\hline\hline
\backslashbox{Matrix}{Index} & $j-1$ & $j$ & $j+1$ \\ [0.5ex]
\hline
$j^{th}\;\text{row of}\;M_{000}$ & $\frac{3}{70}h_{j+1}$ & $\frac{20}{70}(h_j+h_{j+1})$ & $\frac{3}{70}h_{j+1}$ \\
\hline
$j^{th}\;\text{row of}\;M_{100}$ & $\frac{17}{70}$ & $0$ &$-\frac{17}{70}$\\
 [1ex]
\hline
\hline
\end{tabular}
\end{table}

By Substituting the values of $\sigma_{ji}$ and $\eta_{ji}$ from Table 2 in (\ref{80}) and using the definition of $\vartheta_{ji}$, we get the desired consequence.\hfill{$\blacksquare$}
\section{Results and discussions}
Since the main goal of this work is applying the $hp$-cloud method with the stability scheme, most of the discussion (all figures and tables except Table 7) provided here will be about the main stability parameter $\tau_j$ in (\ref{100}) given in Theorem 1. However, only Table 7 sheds some light on the FEM stability parameter given by Lemma 2. This discussion takes a form of comparison with the main stability parameter.

For point nucleus, the relativistic formula is used to compare our results
\begin{equation}\label{91}
\lambda_{n_r,\kappa}=\frac{mc^2}{\sqrt{1+\frac{Z^2\alpha^2} {(n_r-1+\sqrt{\kappa^2-Z^2\alpha^2})^2}}}\,,
\end{equation}
where $\alpha$ is the fine structure constant which has, in atomic unit, the value $1/c$, and the orbital level number $n_r$ takes the values $1,2,\ldots$. To ease performing the comparison, the exact eigenvalues $\lambda_{n_r,\kappa}$ and the positive computed eigenvalues are shifted by $-mc^2$. All computations are performed for the Hydrogen-like Ununoctium ion, where the atomic number and atomic weight for the Ununoctium element are 118 and 294 respectively. Consequently, and since the electron in the Hydrogen-like Ununoctium ion admits relatively large magnitude eigenvalues, for better measuring of the approximation accuracy, through out all computations we shall use the relative error. To treat the singularity of the pure Coulomb potential at $x=0$, extended nucleus is assumed by modifying the potential to fit the finite nuclear size. The modified Coulomb potential considers another distribution of the charge along the nucleus (in the region $[0\,,\,R]$ where $R$ is the nucleus radius) and pure Coulomb potential in the rest of domain. The continuity and the smoothness property (at least $C^1$) should be saved for the total modified potential. For the distribution of charge along nucleus, one can consider, e.g., uniform or Fermi distributions, in this work we consider uniformly distributed charge.

As for the boundary conditions, the homogeneous Dirichlet condition is assumed. Note that for better approximation of the eigenstates $1s_{1/2}$ and $2p_{1/2}$, suitable Neumann boundary conditions should also be considered, see \cite{ALMA}. However, here, we do not treat these cases, instead, general computations are performed to account for the essence of discussion. The homogeneous Dirichlet boundary condition is then simply implemented, after coupling with the FEM, by omitting the two finite element functions at the lower and upper boundaries.

As mentioned before, the computation of the radial Dirac operator eigenvalues requires exponential distribution of the nodal points in order to capture desired behavior of the radial functions near the origin. For this purpose, we shall use the following formula
\begin{equation}\label{90}
x_i=\displaystyle\exp\Big({\ln(I_a+\eps)+\big(\frac{\ln(I_b+\eps)- \ln(I_a+\eps)}{n}\big)i}\Big)-\eps\, , \;\;\; i=0,1,2,\ldots,n,
\end{equation}
where $n$ is the total number of nodal points and $\eps\in[0\,,\,1]$ is the nodes intensity parameter. The role of $\eps$ is to control the intensity of the nodal points close to origin, as smaller $\eps$ as more nodes are dragged toward the origin, see the discussion below. As for other approximation methods, increasing the number of nodal points provides better approximation, but this, of course, on the account of the computational time. However, we can still obtain a good approximation with relatively less time, compared with increasing the nodal points, if the number of integration points is increased (the same size of the generalized matrices is obtained for a fixed number of nodal points, where increasing the number of integration points means more time is needed for functional evaluations but the same time is used for eigenvalues computation of the generalized system). This does not mean that we do not need to increase the number of nodal points to obtain more computed eigenvalues and to improve the approximation, but to get a better rate of convergence with less computational time, increasing both the numbers of integration and nodal points are necessary. In this computation, we fix the number of integration points at $10\cdot n$.

Table 3 shows the approximated eigenvalues of the electron in the Hydrogen-like Ununoctium ion obtained using the usual and the stabilized $hp$-cloud methods, the computation is obtained at $\rho_j=2.2h_{j+1}$, $\eps=10^{-5}$, and $n=600$. The clouds are enriched by $P^t(x)=[1\,,\,x(1-x/2)\,\exp(-x/2)]$. The eigenvalues, through out the computations in this work, are given in atomic unit. In Table 3, with the usual $hp$-cloud method, the instilled spurious eigenvalues appear for both positive and negative $\kappa$ (the two shaded values in the fourteenth level), also the unphysical coincidence phenomenon occurs for the positive $\kappa$ (the shaded value in the first level). Note that these spuriosity of both categories are removed by the $hp$-CPG method.

\begin{table}[h]
\begin{footnotesize}
\caption{The first computed eigenvalues of the electron in the Hydrogen-like Ununoctium ion using the usual and the stabilized $hp$-cloud methods for point nucleus.}
\centering
\begin{tabular}{@{} l c c c c r @{} }
\hline\hline
Level & $hp$-cloud & $hp$-cloud & Exact solution &  $hp$-CPG & $hp$-CPG\\
& $\kappa=2 $ & $\kappa=-2$ & $\kappa=-2$ & $\kappa=-2$ & $\kappa=2$ \\ [0.5ex]
\hline
1&\cellcolor[gray]{0.6}-1829.630750899&-1829.630750902&-1829.630750908&-1829.628309112&\\
2&-826.7698136330&-826.7698136329&-826.7683539069&-826.7714785272&-826.7738882959\\
3&-463.1214970564&-463.1214970566&-463.1183252634&-463.1247150569&-463.1261170024\\
4&-294.4552367950&-294.4552367952&-294.4509801141&-294.4591541031&-294.4600671778\\
5&-203.2468937049&-203.2468937047&-203.2419549027&-203.2511517040&-203.2517946674\\
6&-148.5588260984&-148.5588260983&-148.5534402360&-148.5632453116&-148.5637243357\\
7&-113.2536099083&-113.2536099084&-113.2479180697&-113.2580871797&-113.2584595495\\
8&-89.16385480233&-89.16385480237&-89.15794547564&-89.16832365853&-89.16862284813\\
9&-72.00453396071&-72.00453396065&-71.99846504808&-72.00894720487&-72.00919403005\\
10&-59.35481340095&-59.35481340100&-59.34862423729&-59.35913470352&-59.35934276227\\
11&-49.76429096817&-49.76429096819&-49.75800915710&-49.76849047005&-49.76866900765\\
12&-42.32147184311&-42.32147184312&-42.31511730902&-42.32552373918&-42.32567925216\\
13&-36.43039621976&-36.43039621984&-36.42398370073&-36.43427738957&-36.43441456989\\
14&\cellcolor[gray]{0.6}-33.96502895994&\cellcolor[gray]{0.6}-33.96502895893 &-31.68173025393&-31.69187884728&-31.69200116063\\
15&-31.68818961940&-31.68818961935&-27.80813459180&-27.81810976712&-27.81821982418\\
[1ex]
\hline\hline
\end{tabular}
\label{table:nonlin}
\end{footnotesize}
\end{table}
\subsection{Integration of $hp$-cloud functions\textcolor{white}{.}\\\\}

To approximate the integrals in the weak form in the Galerkin hp-cloud approximation, we use two-point Gaussian quadrature rule. Gaussian quadrature rules are the most used numerical techniques to evaluate the integrals in MMs due to their exactness property in integrating of polynomials of degree $2m_q-1$, where $m_q$ is the number of quadrature points \cite{FRI2}. However, using Gaussian quadrature rules yields integration error when the grids are not coincident with the clouds covers, and thus instabilities and spurious modes start to appear. Also for non-uniformly distributed points (the case we assume in this work), Gaussian quadrature rules do not pass the patch test (fail in consistency). Therefore, stabilizing conforming nodal integration (SCNI), see \cite{CHE1}, is introduced to overcome these difficulties. The main feature of SCNI is using the divergence theorem to substitute the derivative, i.e., the derivative $\frac{d}{dx}\Psi^h$ in the sub-domain $\Omega_j=[x_j\,,\,x_{j+1}]$ is replaced by a smooth derivative (averaging derivative) $\frac{\overline{d}}{dx}\Psi^h$ at $\hat{x}\in\Omega_j$ as
$$
\frac{d}{dx}\Psi^h(x)\approx\frac{\overline{d}}{dx}\Psi^h(\hat{x})= \frac{1}{x_{j+1}-x_j}\displaystyle \int_{x_j}^{x_{j+1}}\frac{d}{dx}\Psi^h(x) dx=\frac{\Psi^h(x_{j+1})-\Psi^h(x_j)}{x_{j+1}-x_j}.
$$
This definition helps stabilizing the integration, further, it saves time in the computation by not calculating the derivatives of the cloud functions. Thus, there is no need to evaluate $(M^{-1})'=-M^{-1}M'M^{-1}$, which is expensive to calculate. For integrating and programming the weak form in MMs, the results from \cite{DOL1,DOL2} are useful.

The cloud shape functions are evaluated at the integration points (digital evaluation), since , practically, it is somehow impossible to write the cloud functions explicitly without matrix inversion. Also, it is not recommended to obtain the inverse of $M$ directly, instead, LU factorization is better to be used from cost (less time consumption) and numerical stability point of views. Moreover, in MMs generally, to enhance the stability of the computation and to maintain the accuracy (that may be affected or lost due to the round-off error), and to get better conditioning of the matrix $M$ (lower condition number), the origin should be shifted to the evaluation point \cite{FRI2,HUE2,LE}, i.e., $x$ is replaced by the transformation $\overline{x}=x-x_{orig}$, consequently $\psi_i(x)=P^t(0)M^{-1}(x)B_i(x)$ where $M(x)\!=\!\displaystyle\sum_{i=1}^n\varphi_i(\frac{x-x_i}{\rho_i})P(x_i-x_{orig})
P^t(x_i-x_{orig})$ and $B_i(x)=\varphi_i(\frac{x-x_i}{\rho_i})P(x_i-x_{orig})$.

\subsection{Enrichment basis functions $P(x)$\textcolor{white}{.}\\\\}

For the reason discussed before, only intrinsic enrichment, $P(x)$, is considered in the definition of the $hp$-cloud functions for the computation of the eigenvalues of the radial Dirac operator. The number and the type of enrichment functions in the basis set $P(x)$ can be chosen arbitrary for each cloud \cite{GAR,MEN}, but for practical reasons (lowering both the condition number of $M$ and the computational time) we assume $P(x)=[1, p_1(x)]$. For the approximation of the radial Dirac operator eigenvalues, to enrich the cloud with a suitable basis $P(x)$, two main properties should be considered; firstly, and sufficient one, the elements of $P(x)$ ought to have the continuity properties (continuous with continuous first derivatives) of the space $\mathcal{H}$ so that for all $j$, the cloud $\psi_j$ is a $C^1$-function, provided that $\varphi_j$ is a $C^1$-function. Secondly, global behavior and fundamental characters about the electron motion of the Hydrogen-like ion systems should be embedded in $P(x)$. Slater type orbital functions (STOs) and Gaussian type orbital functions (GTOs) provide good description of the electron motion \cite{CHE2,HEH}. The quadratic term in the exponent of the GTOs causes numerical difficulty, that is, with the GTOs, the matrix $M$ rapidly becomes poorly conditioned, this is also what is observed when applying quadratic basis enrichments, see \cite{BEL2}. Consequently, the STOs are considered as the enrichment of the $hp$-cloud functions, thus $p_1(x)$ can have, e.g., the following forms
$$
\exp(-x),\;x\,\exp(-x/2),\;x(1-x/2)\,\exp(-x/2),\ldots\; \text{etc.}
$$
Note that, these functions should be multiplied by normalization parameters, but, computationally, there is no effect of multiplication by these parameters.

Since the global behavior of the eigenstates of the Hydrogen-like ions in the relativistic case (Dirac operator) does not differ much from that of the non-relativistic case (Schr\"odinger operator), one can also assume the solutions of the radial Coulomb-Schr\"odinger eigenvalue problem as intrinsic enrichments (see e.g. \cite{HIL})
$$
\mathcal{R}_{n_r\ell}(x)=\mathcal{N}_{n_r\ell}\,(2Zx/n_ra_0)^\ell\; \displaystyle\mathbf{L}_{n_r+\ell}^{2\ell+1}(2Zx/n_ra_0)\, \exp(-Zx/n_ra_0),
$$
where
$\displaystyle\mathbf{L}_{n_r+\ell}^{2\ell+1}(x)=\displaystyle\sum_{k=0}^{n_r+ \ell}\frac{(-1)^k}{k!}\left(
\begin{array}{c}
n_r+3\ell+1 \\
n_r+\ell-k
\end{array}
\right)
x^k$ is the Laguerre polynomial, $a_0$ is the Bohr radius, $n_r=1,2,\ldots$ is the orbital level number, and $\ell$ is, as mentioned before, the orbital angular momentum number given to be zero for $s$-states, one for $p$-states, two for $d$-states, etc. For general intrinsic enrichment, it is, somehow, tedious to apply the above formula for each level $n_r$, instead, good results are still achievable even with, e.g., $n_r$ equals the first possible level of the given state (i.e., $n_r=1$ for all $s$-states, $n_r=2$ for all $p$-states, $n_r=3$ for all $d$-states, etc.). Moreover, it is also possible to consider enrichment based on the solution of the radial Dirac eigenvalue problem, see e.g. \cite{CIF}, but the above enrichments are simpler from practical point of view. In the coming discussion, the enrichment basis $P^t(x)=[1\,,\,x(1-x/2)\,\exp(-x/2)]$ is assumed in all computations.

\subsection{Dilation Parameter $\rho$\textcolor{white}{.}\\\\}

The dilation parameter, $\rho$, plays a crucial role in the approximation accuracy and stability, it serves as the element size in the FEM. The parameter $\rho$ can be chosen fixed or arbitrary, but it is often assumed to be constant for all $hp$-clouds when uniformly distributed nodal points are used. In this work, exponentially distributed nodal points are assumed to get enough information about the radial functions near the origin where they oscillate heavily relative to a region away from it. Thus we consider
$$
\rho_j=\nu\cdot max\{h_j\,,\,h_{j+1}\}=\nu h_{j+1},
$$
where the maximum is considered to engage sufficiently large region where the cloud function is defined so that the risk for singularity of the matrix $M$ is substantially decreased, further, $\nu$ is the dimensionless size of influence domain \cite{LE}. Moreover, for a non-uniform mesh, $\nu$ can be chosen locally, i.e., $\nu=\nu_j$, where, in this work, we assume fixed $\nu$. Now it remains to determine the value$/$values of $\nu$ taking into account that $\rho_j$ should be large enough ($\nu>1$) in order to ensure the invertibility of $M$ (to ensure that any region is covered by at least two clouds). On the other hand, $\rho_j$ should not be very large to maintain local character of the approximation. As discussed before (see also Figure 3, the case $\nu=1.2$), if $\nu\rightarrow1$, then $\psi_j$ will act as finite element shape function, and thus the features of the $hp$-clouds are gradually lost, also large values of $\nu$ make $\psi_j$ to behave like polynomial of higher degree (see Figure 3, the case $\nu=4$). To conclude, $\nu$ should be chosen moderately and such that it guarantees that no integration point is covered by only one cloud \cite{LE,MEN}.

The optimal choices of $\nu$ can be determined individually for each problem by carrying out numerical experiments. In \cite{LIU,YOU}, it is shown that $\nu\in[2,3]$ provides nice results for the elasticity problem. For the computation of the radial Dirac operator eigenvalues with the stability scheme, when $\nu\in[2.2,2.7]$ good results are obtained and the spurious eigenvalues are completely eliminated. Also as $\nu$ gets smaller as better approximation is obtained, see Table 4.
\begin{table}[h]
\begin{footnotesize}
\caption{The first computed eigenvalues of the electron in the Hydrogen-like Ununoctium ion for $\kappa=-2$ for point nucleus with different values of $\nu$, where $n=600$ and $\eps=10^{-5}$ are used.}
\centering
\begin{tabular}{@{} l c c c c r @{} }
\hline\hline
Level & $\nu=2.0$ & $\nu=2.2$ & $\nu=2.5$ &  $\nu=2.7$ & Exact solution \\ [0.5ex]
\hline
1&-1829.6287&-1829.6283&-1829.6276&-1829.6270&-1829.6307\\
2&-826.77119&-826.77147&-826.77197&-826.77233&-826.76835\\
3&-463.12417&-463.12471&-463.12567&-463.12638&-463.11832\\
4&-294.45850&-294.45915&-294.46033&-294.46120&-294.45098\\
5&-203.25046&-203.25115&-203.25244&-203.25340&-203.24195\\
6&-148.56255&-148.56324&-148.56460&-148.56562&-148.55344\\
7&-113.25741&-113.25808&-113.25949&-113.26054&-113.24791\\
8&-89.167688&-89.168323&-89.169756&-89.170831&-89.157945\\
9&-72.008358&-72.008947&-72.010396&-72.011489&-71.998465\\
10&-59.358602&-59.359134&-59.360592&-59.361700&-59.348624\\
11&-49.768025&-49.768490&-49.769950&-49.771070&-49.758009\\
12&-42.325133&-42.325523&-42.326981&-42.328113&-42.315117\\
13&-36.433970&-36.434277&-36.435728&-36.436870&-36.423983\\
14&-31.691663&-31.691878&-31.693318&-31.694472&-31.681730\\
15&-27.817992&-27.818109&-27.819533&-27.820699&-27.808134\\ [1ex]
\hline\hline
\end{tabular}
\label{table:nonlin}
\end{footnotesize}
\end{table}
In Figure 4, we study the convergence rate of the first five eigenvalues in Table 4.
\begin{figure}[H]
\centering
\includegraphics[width=7.8cm]{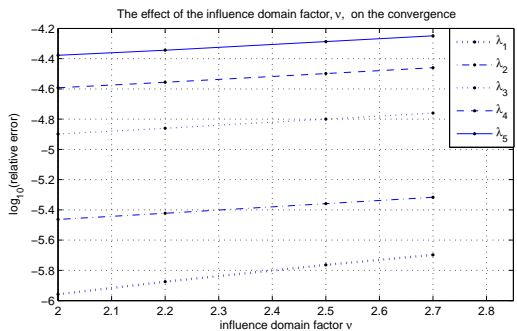}
\caption{Studying the convergence rate with respect to the influence domain factor $\nu$. The comparison is carried out for the first five eigenvalues in Table 4.}
\end{figure}
It is clear how the smaller $\nu$ gives the better approximation. One argues, as it is clear from the figure, that $\nu$ can be, e.g., of some value less than 2 in order to achieve a better rate of convergence. However, this will be on the account of spuriosity elimination (the cloud is not stretched enough to capture the desired behavior of the approximated solution) and on the account of the invertibility of the matrix $M$ (for small $\nu$ some regions are covered with one cloud). However, as in the FEM, one can apply $h$-refinement in the $hp$-cloud method (see e.g. \cite{DUA2,YOU}), this can be done by assuming smaller values of the dilation parameter $\rho_j$ (keeping $\nu$ fixed and making $h_{j+1}$ smaller by increasing the number of nodal points). Thus as $\rho_j$ getting smaller, more clouds of smaller domain sizes are added.\\

The intensity of the exponentially distributed nodal points near the origin has an influence on the convergence rate of the approximation. The intensity of the nodes, near the origin or away from it, is controlled by the nodes intensity parameter, $\eps$, via formula (\ref{90}). As smaller value of $\eps$ is considered as more concentration of nodes near the origin is obtained, see Figure 5 (the graph to the left).

Table 5 shows the computation of the eigenvalues with different values of $\eps$ with $600$ nodal points. The computation with $\eps$ smaller than $10^{-7}$ is almost the same as of $\eps=10^{-7}$, thus it is not required to study smaller values of $\eps$ than $10^{-7}$.
\begin{table}[h]
\begin{footnotesize}
\caption{The first computed eigenvalues of the electron in the Hydrogen-like Ununoctium ion for $\kappa=-2$ for point nucleus with different values of $\eps$, where $n=600$ and $\nu=2.2$ are used.}
\centering
\begin{tabular}{@{} l c c c c r @{} }
\hline\hline
Level & $\eps=10^{-4}$ & $\eps=10^{-5}$ & $\eps=10^{-6}$ &  $\eps=10^{-7}$ & Exact solution \\ [0.5ex]
\hline
1&-1829.6289&-1829.6283&-1829.6280&-1829.6280&-1829.6307\\
2&-826.77073&-826.77147&-826.77170&-826.77173&-826.76835\\
3&-463.12322&-463.12471&-463.12517&-463.12523&-463.11832\\
4&-294.45726&-294.45915&-294.45973&-294.45981&-294.45098\\
5&-203.24904&-203.25115&-203.25180&-203.25188&-203.24195\\
6&-148.56101&-148.56324&-148.56393&-148.56402&-148.55344\\
7&-113.25578&-113.25808&-113.25879&-113.25888&-113.24791\\
8&-89.165992&-89.168323&-89.169039&-89.169131&-89.157945\\
9&-72.006610&-72.008947&-72.009662&-72.009755&-71.998465\\
10&-59.356811&-59.359134&-59.359844&-59.359936&-59.348624\\
11&-49.766195&-49.768490&-49.769189&-49.769279&-49.758009\\
12&-42.323268&-42.325523&-42.326208&-42.326296&-42.315117\\
13&-36.432073&-36.434277&-36.434943&-36.435030&-36.423983\\
14&-31.689734&-31.691878&-31.692524&-31.692607&-31.681730\\
15&-27.816033&-27.818109&-27.818732&-27.818812&-27.808134\\
[1ex]
\hline\hline
\end{tabular}
\label{table:nonlin}
\end{footnotesize}
\end{table}

In Figure 5 (the graph to the right), the first computed eigenvalues of Table 5 are studied. It is clear that as $\eps$ gets larger (up to some limit), better approximation is obtained. However, as mentioned before, the rate of convergence is almost the same when $\eps\in(0\,,\,10^{-7})$ ($\eps=0$ is excluded to avoid $\ln(0)$ when extended nucleus is assumed). Also $\eps$ does not admit relatively large values in order to get enough nodes close to the origin, where the radial functions oscillate relatively more, without increasing the number of nodal points. Therefore, the most appropriate values for $\eps$ which provide good results, are somewhere in the interval $[10^{-6}\,,\,10^{-4}]$.

\begin{figure}[h]
\centering
\includegraphics[width=7.8cm]{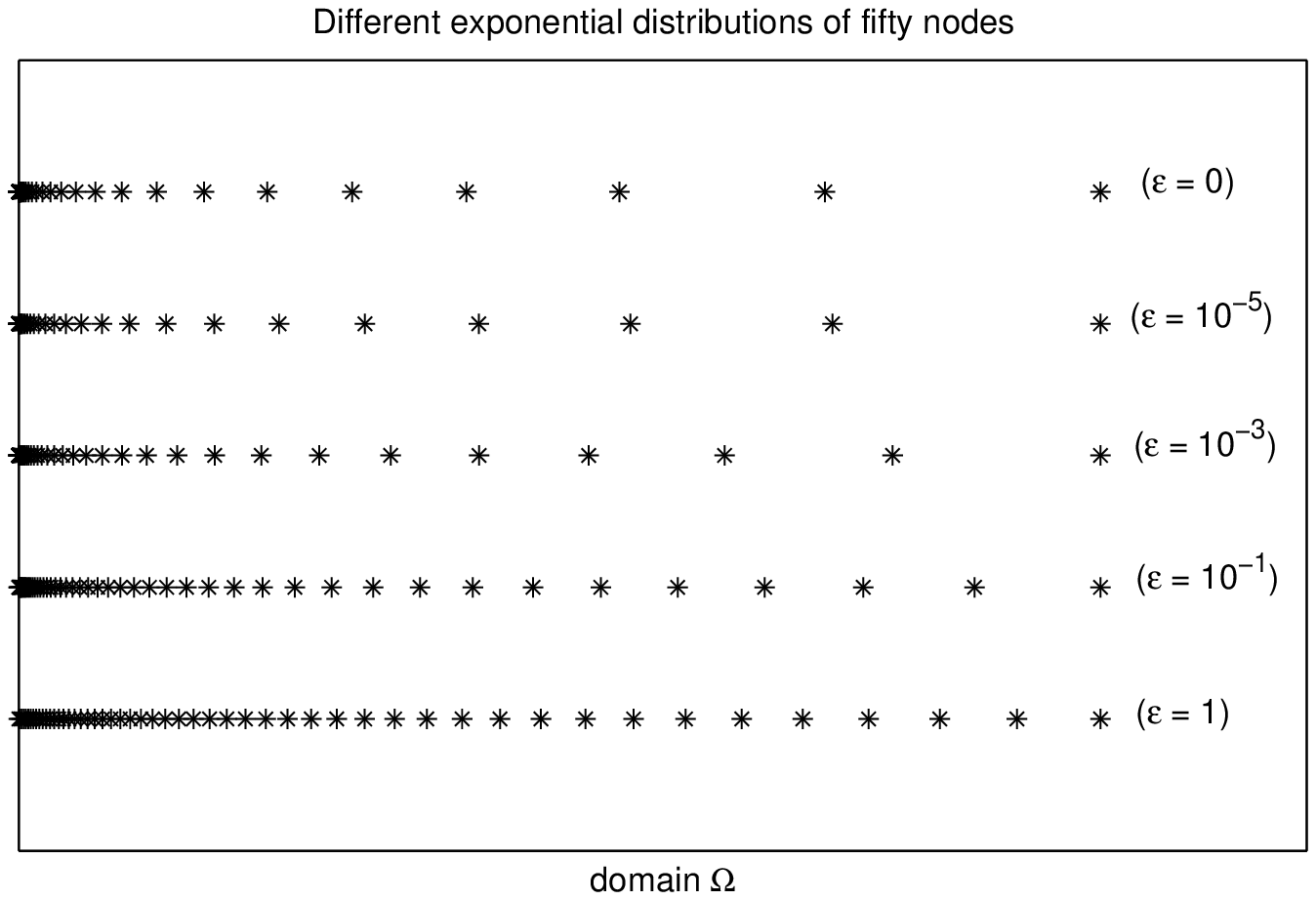}
\includegraphics[width=7.8cm]{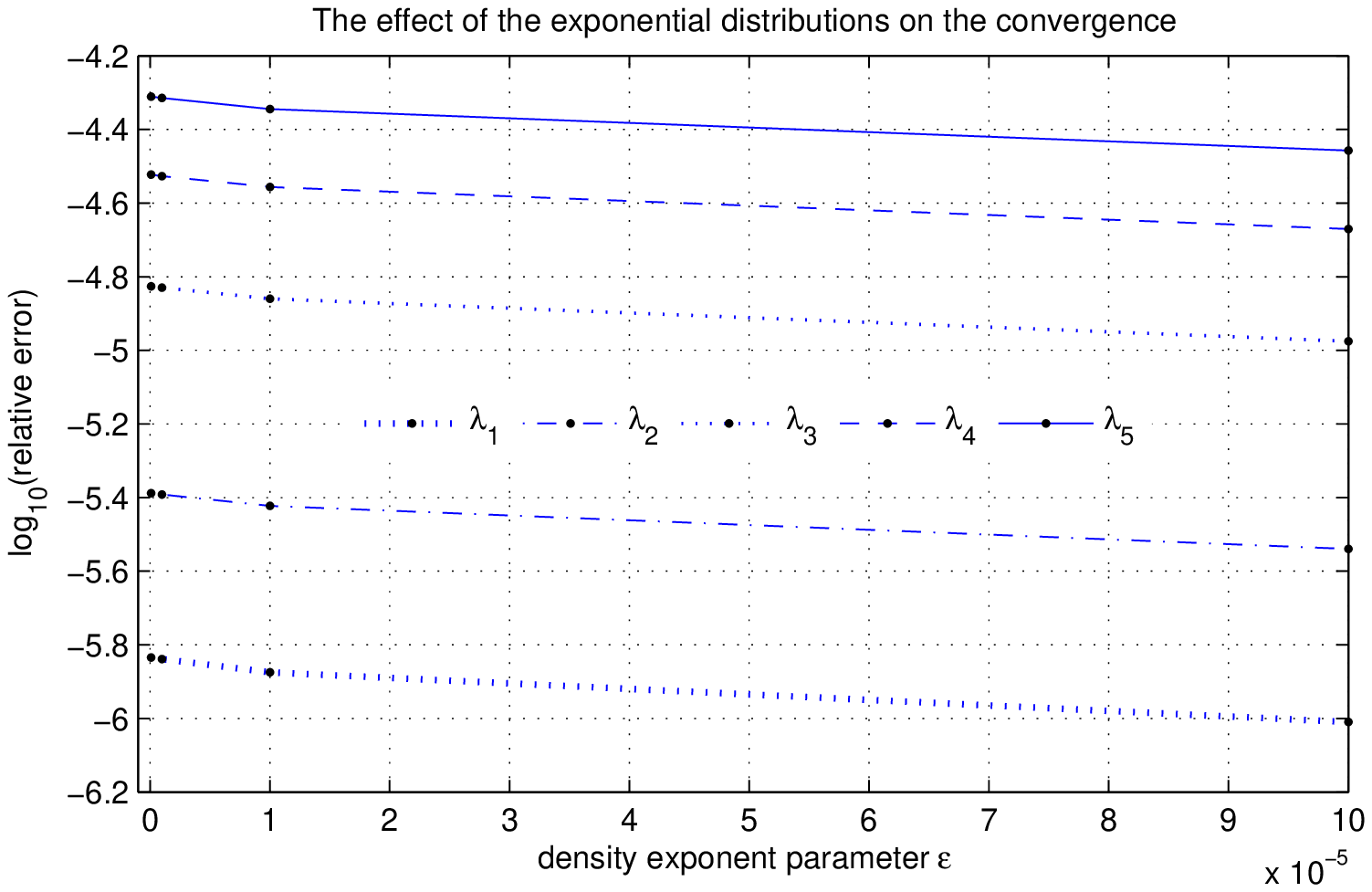}
\caption{To the left, different exponentially distributed nodal points are plotted using the formula (\ref{90}). To the right, the effect of nodes intensity near the origin on the convergence rate, the comparison is carried out for the first five eigenvalues in Table 5.}
\end{figure}

The approximation of the stabilized $hp$-cloud scheme with different numbers of nodal points is given in Table 6. The rate of convergence of the corresponding first five eigenvalues is studied in Figure 6, where $h$ is the maximum of all distances between the adjacent nodes which equals to $h_n=x_n-x_{n-1}$, the distance between the last two nodes for exponentially distributed nodes.

\begin{table}[h]
\begin{footnotesize}
\caption{The first computed eigenvalues of the electron in the Hydrogen-like Ununoctium ion for $\kappa=-2$ for point nucleus with different number of nodes, where $\nu=2.2$ and $\eps=10^{-5}$ are used.}
\centering
\begin{tabular}{@{} l c c c c c r @{} }
\hline\hline
Level & $n=200$ & $n=400$ & $n=600$ &  $n=800$ &  $n=1000$ & Exact solution \\ [0.5ex]
\hline
1 &-1829.5628&-1829.6224&-1829.6283&-1829.6297&-1829.6302&-1829.6307\\
2 &-826.82670&-826.77726&-826.77147&-826.76987&-826.76923&-826.76835\\
3 &-463.23292&-463.13630&-463.12471&-463.12146&-463.12016&-463.11832\\
4 &-294.59147&-294.47367&-294.45915&-294.45503&-294.45336&-294.45098\\
5 &-203.39386&-203.26721&-203.25115&-203.24654&-203.24466&-203.24195\\
6 &-148.70878&-148.58009&-148.56324&-148.55835&-148.55635&-148.55344\\
7 &-113.40170&-113.27527&-113.25808&-113.25304&-113.25096&-113.24791\\
8 &-89.306709&-89.185557&-89.168323&-89.163201&-89.161076&-89.157945\\
9 &-72.139617&-72.026008&-72.008947&-72.003802&-72.001653&-71.998465\\
10&-59.480154&-59.375861&-59.359134&-59.354006&-59.351849&-59.348624\\
11&-49.878353&-49.784751&-49.768490&-49.763410&-49.761256&-49.758009\\
12&-42.423104&-42.341207&-42.325523&-42.320517&-42.318374&-42.315117\\
13&-36.518814&-36.449288&-36.434277&-36.429365&-36.427242&-36.423983\\
14&-31.762955&-31.706134&-31.691878&-31.687081&-31.684984&-31.681730\\
15&-27.875610&-27.831538&-27.818109&-27.813442&-27.811376&-27.808134\\
[1ex]
\hline\hline
\end{tabular}
\label{table:nonlin}
\end{footnotesize}
\end{table}

The lack of error estimates for the approximation of the Dirac eigenvalue problem due to the boundedness problem results in an incomplete picture about the convergence analysis. Nevertheless, from Figure 6, the convergence rate of the approximation of the first five eigenvalues, $\lambda_1$, $\ldots$, $\lambda_5$, is nearly $3.09$, $2.66$, $2.62$, $2.59$, and $2.56$ respectively, which it takes a slight decreasing pattern as we go higher in the spectrum levels, see the corresponding table.\\\\

\begin{figure}[h]
\centering
\includegraphics[width=7.8cm]{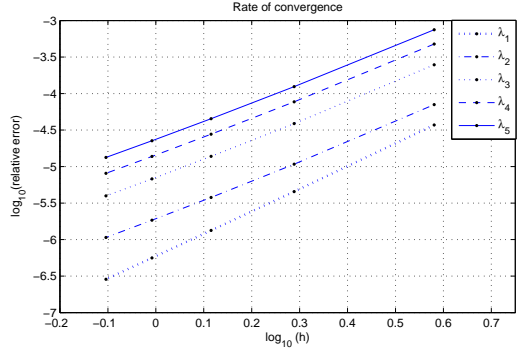}
\caption{Study the convergence rate of the first computed five eigenvalues in Table 6.}
\end{figure}

With the stability parameter $\tau^{F\!E\!M}$, the computation is presented in Table 7. The computation is obtained with 600 nodal points at $\nu=1.1$ and $\eps=10^{-5}$. The result is compared with the same stability scheme but with the stability parameter $\tau$ at the same parameters but $\nu=2.2$, the comparison is also obtained in the non-relativistic limit (very large $c$).
\begin{table}[h]
\begin{footnotesize}
\caption{The first computed eigenvalues of the electron in the Hydrogen-like Ununoctium ion for $\kappa=-2$ for point nucleus using the stability scheme with the stability parameters $\tau$ and $\tau^{F\!E\!M}$.}
\centering
\begin{tabular}{@{} l c c c|| c c r @{} }
\hline\hline
&& The speed of light & & & $100\times$The speed of light &  \\
Level&$\tau$ & $\tau^{F\!E\!M}$ & Exact solution & $\tau$ & $\tau^{F\!E\!M}$ & Exact solution \\ [0.5ex]
\hline
1&-1829.6283&-1829.6304&-1829.6307&-1740.2372&-1740.4777&-1740.5080\\
2&-826.77147&-826.76993&-826.76835&-773.73860&-773.57259&-773.56033\\
3&-463.12471&-463.12174&-463.11832&-435.46054&-435.14787&-435.12752\\
4&-294.45915&-294.45551&-294.45098&-278.88245&-278.49775&-278.48144\\
5&-203.25115&-203.24715&-203.24195&-193.82362&-193.39522&-193.38978\\
6&-148.56324&-148.55905&-148.55344&-142.53145&-142.07261&-142.08222\\
7&-113.25808&-113.25377&-113.24791&-109.23625&-108.75140&-108.78165\\
8&-89.168323&-89.163916&-89.157945&-86.404375&-85.894382&-85.950912\\
9&-72.008947&-72.004478&-71.998465&-70.067886&-69.534118&-69.620219\\
10&-59.359134&-59.354644&-59.348624&-57.975599&-57.420335&-57.537357\\
11&-49.768490&-49.764010&-49.758009&-48.773149&-48.197640&-48.347352\\
12&-42.325523&-42.321064&-42.315117&-41.606088&-41.009232&-41.195370\\
13&-36.434277&-36.429826&-36.423983&-35.913753&-35.292096&-35.520492\\
14&-31.691878&-31.687405&-31.681730&-31.315908&-30.664671&-30.942291\\
15&-27.818109&-27.813579&-27.808134&-27.547311&-26.861865&-27.195369\\
[1ex]
\hline
\hline
\end{tabular}
\label{table:nonlin}
\end{footnotesize}
\end{table}

As it can be seen from Table 7, the convergence property with $\tau^{F\!E\!M}$ is slightly better. Unfortunately, the approximation with $\tau^{F\!E\!M}$ seems to behave strangely at the end of the spectrum, that is, only the spectrum tail has the following behavior (the last eigenvalues of the computation in Table 7 with $\tau^{F\!E\!M}$ for the relativistic case)\\

 \hspace{3.75cm} $\lambda_+ -mc^2\quad\quad\quad\quad\quad$  $\lambda_- +mc^2$
 \begin{verbatim}
                   207072481.0215   -215565247.3448
                   211429663.4158*  -220006205.1800*

                   226003907.3130   -235294474.7992
                   231896256.0483*  -241138935.9851*

                   246890583.9362   -257366374.4374
                   257292411.7094*  -267386241.2969*
                   267659710.2673*  -279193268.7275*

                   291928112.6166   -303237209.5231
                   296228215.8873*  -308029351.9019*
 \end{verbatim}
This behavior occurs only for few values at the end of the spectrum, and no such effect is revealed in the rest of the spectrum. To our knowledge, the values marked with $\boldsymbol{*}$ might be spurious eigenvalues for some unknown origins in higher levels, which, in calculating the correlation energy, seem to have no significant effect.

Table 8 shows the computation of the eigenvalues of the electron in the Hydrogen-like Ununoctium ion with $\kappa=-2$. The computation is for extended nucleus obtained using the stability scheme, where the first and the last computed eigenvalues are presented. The number of nodes used is 1000, also the used values of $\nu$ and $\eps$ are respectively $2.2$ and $10^{-5}$.\\

\textbf{Conclusion.\\}

The scheme developed in this work, the $hp$-CPG method, for stabilizing the $hp$-cloud approximation for solving the single-electron Coulomb-Dirac eigenvalue problem ensures complete treatment of the spurious eigenvalues. The scheme strongly depends on the derived stability parameter $\tau$, which is simple to implement and applicable for general finite basis functions. The elimination of the spurious eigenvalues is also affected by the influence domain factor $\nu$, for $\nu$ less than 2, spurious eigenvalues start to appear. The convergence rate is high for the first eigenvalues, while it slowly decreases as the level gets higher. Comparing with the finite element stability approach \cite{ALMA}, the scheme convergence rate is lower. We may state that, as the main disadvantage of MMs in general, the $hp$-cloud method is more expensive due to the time consumption in evaluating the shape function which demands more integration point as $\nu$ gets larger to obtain the desired accuracy. The number of integration points used here is ten times the number of nodal points (this large number of points is assumed in order to study the effects of the other parameters from a comparative point of view), which can be made smaller, i.e., $\nu\geq2$ is enough to get sufficient accuracy.

\begin{table}[H]
\begin{footnotesize}
\caption{The first and the last computed eigenvalues of the electron in the Hydrogen-like Ununoctium ion for $\kappa=-2$ for extended nucleus using the stability scheme.}
\centering
\begin{tabular}{@{} l c c|| c c r @{} }
\hline\hline
Level&$\lambda_+ -mc^2$ & $\lambda_- +mc^2$ & Level &$\lambda_+ -mc^2$ & $\lambda_- +mc^2$ \\ [0.5ex]
\hline
1&-1829.630099296&-2.434417024833&956&586688854.9879&-592440657.5171\\
2&-826.7693836687&-2.627617735663&957&598438233.4986&-604235496.3134\\
3&-463.1204703095&-2.797844649762&958&610919226.3899&-616906339.6739\\
4&-294.4537598806&-2.957820323769&959&623291095.2201&-629316989.3797\\
5&-203.2451112716&-3.112227360954&960&636480181.4813&-642726428.9931\\
6&-148.5568310625&-3.263408080529&961&649524826.4034&-655799753.0246\\
7&-113.2514598678&-3.412742089389&962&663483591.3527&-670016495.9307\\
8&-89.16158600002&-3.561138820860&963&677258040.9295&-683804920.4801\\
9&-72.00216968256&-3.709254718702&964&692056051.2167&-698907640.1176\\
10&-59.35236884621&-3.857601123902&965&706625519.7561&-713470210.1966\\
11&-49.76177611950&-4.006600112632&966&722341703.9586&-729549529.5797\\
12&-42.31889320986&-4.156612122006&967&737781471.5140&-744953276.6295\\
13&-36.42775793119&-4.307947863723&968&754505800.5837&-762114239.5911\\
14&-31.68549413213&-4.460872692773&969&770903696.8948&-778435989.9532\\
15&-27.81188079379&-4.615608997468&970&788739250.6668&-796801160.7705\\
16&-24.60715078621&-4.772339706105&971&806198985.1845&-814129982.6368\\
17&-21.92579810728&-4.931213787881&972&825264528.2332&-833843359.2405\\
18&-19.65981495410&-5.092353196094&973&843910220.2386&-852283956.3394\\
19&-17.72767091768&-5.255860198286&974&864343472.0411&-873515941.8428\\
20&-16.06689248232&-5.421824189525&975&884325905.7234&-893193487.1967\\
21&-14.62895927631&-5.590327472124&976&906287797.8413&-916147241.8477\\
22&-13.37572611330&-5.761449842514&977&927793189.6390&-937214464.1559\\
23&-12.27687110094&-5.935272041930&978&951473602.6651&-962134060.6197\\
24&-11.30804689661&-6.111878218067&979&974736078.1642&-984781962.7117\\
25&-10.44952215915&-6.291357562437&980&1000361933.873&-1011962889.837\\
26&-9.685170122048&-6.473805266356&981&1025681563.220&-1036437526.155\\
27&-9.001706528736&-6.659322921776&982&1053528920.945&-1066240230.486\\
28&-8.388109077412&-6.848018469445&983&1081298213.911&-1092869973.210\\
29&-7.835170608265&-7.040005799103&984&1111711637.956&-1125737288.414\\
30&-7.335151949937&-7.235404092542&985&1142455141.979&-1154979024.500\\
31&-6.881509811941&-7.434337004637&986&1175881200.913&-1191458619.345\\
32&-6.468681741462&-7.636931764649&987&1210315776.667&-1223979724.299\\
33&-6.091914867255&-7.843318267751&988&1247366017.754&-1264753730.099\\
34&-5.747128531017&-8.053628212296&989&1286494208.919&-1301585190.310\\
35&-5.430803353698&-8.267994317084&990&1328074902.403&-1347514563.183\\
36&-5.139891080678&-8.486549638106&991&1373330615.013&-1390354085.681\\
37&-4.871740872106&-8.709426988262&992&1420940894.294&-1442573015.641\\
38&-4.624038699727&-8.936758453401&993&1474405770.317&-1494428385.378\\
39&-4.394757252718&-9.168674990411&994&1530930196.971&-1554661710.974\\
40&-4.182114320607&-9.405306088338&995&1595537987.175&-1621361454.883\\
41&-3.984538053675&-9.646779469269&996&1667868144.599&-1693349701.469\\
42&-3.800637833043&-9.893220807513&997&1746516931.180&-1787808061.719\\
43&-3.629179737774&-10.14475343997&998&1858171146.415&-1885927735.358\\
44&-3.469065800899&-10.40149804032&999&1944896072.579&-2040151500.838\\
45&-3.319316399200&-10.66357222412&1000&2551096858.208&-2992548052.333\\
[1ex]
\hline
\hline
\end{tabular}
\label{table:nonlin}
\end{footnotesize}
\end{table}


\begin{thebibliography}{References}\setlength{\itemsep}{1.5mm}
\bibitem{ACK} E. Ackad and M. Horbatsch, \emph{Numerical solution of the Dirac equation by a mapped fourier grid method}, J. Phys. A: Math. Gen., 38(2005), pp. 3157-3171.
\bibitem{ALMA} H. Almanasreh, S. Salomonson, and N. Svanstedt, \emph{Stabilized finite element method for the radial Dirac equation}, J. Comp. Phys., 236(2013), pp. 426-442.
\bibitem{ALME} R. C. Almeida and R. S. Silva, \emph{A stable Petrov-Galerkin method for convection-dominated problems}, Comput. Methods Appl. Mech. Engng., 140(1997).
\bibitem{ATL} S. N. Atluri and S. Shen, \emph{The meshless local Petrov-Galerkin (MLPG) method: A simple and less-costly alternative to the finite element and boundary elemenet methods}, CMES, 3(2002), pp. 11-51.
\bibitem{BEL} T. Belytschko, D. Organ, and Y. Krongauz, \emph{A coupled finite element-element-free Galerkin method}, Comput. Mech., 17(1995), pp. 186-195.
\bibitem{BEL2} T. Belytschko, Y. Krongauz, D. Organ, M. Fleming, and P. Krysl, \emph{Meshless methods: An overview and recent developments}, Comput. Methods Appl. Mech. Engng., 139(1996), pp. 3-47.
\bibitem{BRO81} A. N. Brooks, \emph{A Petrov-Galerkin finite element formulation for convection dominated flows}, Thesis for the degree of Doctor of Philosophy, California Institute of Technology, California, 1981.
\bibitem{BRO82} A. N. Brooks and T. J. R. Hughes, \emph{Streamline Upwind/Petrove-Galerkin formulations for convection dominated flows with particular emphasis on the incompressible Navier-Stokes equations}, Comput. Methods Appl. Mech. Engng., 32(1982).
\bibitem{CHE1} J. S. Chen, C. T. Wu, S. Yoon, and Y. You, \emph{A stabilized conforming nodal integration for Galerkin mesh-free methods}, Int. J. Numer. Methods Engng., 50(2001), pp. 435-466.
\bibitem{CHE2} J. S. Chen, W. Hu, and M. A. Puso, \emph{Orbital $hp$-clouds for solving Schr\"odinger equation in quantum mechanics}, Comput. Methods Appl. Mech. Engng., 196(2007), pp. 3693-3705.
\bibitem{CIF} H. Ciftci, R. L. Hall, and N. Saad, \emph{Iterative solutions to the Dirac equation}, Phys. Rev. A, 72(2005).
\bibitem{DES} P. A. B. De Sampaio, \emph{A Petrov-Galerkin/modified operator formulation for convection-diffusion problems}, Int. J. Numer. methods Engng., 30(1990).
\bibitem{DOL1} J. Dolbow and T. Belytschko, \emph{An introduction to programming the meshless element free Galerkin method}, Arch. Comput. Method E., 5(1998), pp. 207-241.
\bibitem{DOL2} J. Dolbow and T. Belytschko, \emph{Numerical integration of the Galerkin weak form in meshfree methods}, Comput. Mech., 23(1999), pp. 219-230.
\bibitem{DUA} C. A. Duarte and J. T. Oden, \emph{$H$-$p$ clouds---An $h$-$p$ meshless method}, Numer. Meth. Part. D. E., 12(1996), pp. 673-705.
\bibitem{DUA2} C. A. Duarte and J. T. Oden, \emph{An $h$-$p$ adaptive method using clouds}, Comput. Methods Appl. Mech. Engng., 139(1996), pp. 237-262.
\bibitem{FRI1} T. Fries and H. Matthies, \emph{A review of Petrov-Galerkin stabilization approaches and an extension to meshfree methods}, Institiute of scientific computing, Technical University Braunschweig, Brunswick, Germany, 2004.
\bibitem{FRI2} T. Fries and H. Matthies, \emph{Classification and overview of Meshfree Methods}, Institiute of scientific computing, Technical University Braunschweig, Brunswick, Germany, 2004.
\bibitem{GAR} O. Garcia, E. A. Fancello, C. S. de Barcellos, and C. A. Duarte, \emph{$hp$-Clouds in Mindlin's thick plate model}, Int. J. Numer. methods Engng., 47(2000), pp. 1381-1400.
\bibitem{GRI} M. Griesemer and J. Lutgen, \emph{Accumulation of Discrete Eigenvalues of the Radial Dirac Operator}, J. Funct. Anal., 162(1999).
\bibitem{HEH} W. J. Hehre, L. Radom, P. Schleyer, and J. Pople, \emph{Ab initio molecular orbital theory}, John Willey $\&$ Sons, New York, 1986.
\bibitem{HIL} W. T. Hill and C. H. Lee, \emph{Light-Matter interaction: Atoms and molecules in external fields and nonlinear optics}, Wiley-VCH Verlag GmbH $\&$ Co. KGaA, Weinheim, 2007.
\bibitem{HUE} A. Huerta and S. Fern\'{a}ndez-M\'{e}ndez, \emph{Coupling element free Galerkin and finite element methods}, ECCOMAS, Barcelona, 2000.
\bibitem{HUE2} A. Huerta and S. Fern\'{a}ndez-M\'{e}ndez, \emph{Enrichment and coupling of the finite element and meshless methods}, Int. J. Numer. methods Engng., 48(2000), pp. 1615-1636.
\bibitem{IDE} S. Idelsohn, N. Nigro, M. Storti, and G. Buscaglia, \emph{A Petrov-Galerkin formulation for advection-reaction-diffusion problems}, Comput. Methods Appl. Mech. Engng., 136(1996).
\bibitem{KRO} Y. Krongauz and T. Belytscko, \emph{Enforcement of essential boundary conditions in meshless approximations using finite elements}, Comput. Methods Appl. Mech. Engng., 131(1996), pp. 133-145.
\bibitem{LE} C. V. Le, \emph{Novel numerical procedures for limit analysis of structures: Mesh-free methods and mathematical programming}, Ph.D thesis, University of Sheffield, England, 2009.
\bibitem{LIN} H. Lin and S. N. Atluri, \emph{Meshless local Petrov-Galerkin (MLPG) method for convection-diffusion problems}, CMES, 1(2000), pp. 45-60.
\bibitem{LIND} I. Lindgren, S. Salomonson, and B. {\AA}s\'{e}n, \emph{The covariant-evolution-operator method in bound-state QED}, Physics Reports, 389(2004), pp. 161-261.
\bibitem{LIU} G. R. Liu, \emph{Mesh Free Methods: Moving Beyond the Finite Element Method}, CRC press, 2003.
\bibitem{LU} Y. Y. Lu, T. Belytschko, and L. Gu, \emph{A new implementation of the element free Galerkin method}, Comput. Methods Appl. Mech. Engng, 113(1994), pp. 397-414.
\bibitem{MEN} P. de T. R. Mendon\c{c}a, C. S. de Barcellos, A. Duarte, \emph{Investigations on the $hp$-cloud method by solving Timoshenko beam problems}, Comput. Mech., 25(2000), pp. 286-295.
 \bibitem{MOH} P. J. Mohr, G. Plunien, and G. Soff, \emph{QED corrections in heavy atoms}, Physics Reports, 293(1998), pp. 227-369.
\bibitem{MUR} G. Mur, \emph{On the causes of spurious solutions in electromagnetics}, Electromagnetic, 22(2002), pp. 357-367.
\bibitem{NGU} V. P. Nguyen, T. Rabczuk, S. Bordas, and M. Duflot, \emph{Meshless methods: A review and computer implementation aspects}, Math. Comput. Simulat., 79(2008), pp. 763-813.
\bibitem{NOG} H. Noguchi, T. Kawashima, and T. Miyamura, \emph{Element free analyses of shell and spatial strucures}, Int. J. Numer. methods Engng., 47(2000), pp. 1215-1240.
\bibitem{PES} G. Pestka, \emph{Spurious roots in the algebraic Dirac equation}, Chem. Phys. Lett. 376(2003), pp. 659-661.
\bibitem{ROS} L. Rosenberg, \emph{Virtual-pair effects in atomic structure theory}, Phys. Rev. A, 39(1989), pp. 4377-4386.
\bibitem{SCH} W. Schroeder and I. Wolf, \emph{The origin of spurious modes in numerical solutions of electromagnetic field eigenvalue problems}, IEEE Tran. on Micr. Theory and Tech., 42(1994), pp. 644-653.
\bibitem{SHAB1} V. M. Shabaev, \emph{Two-time Green's function method in quantum electrodynamics of high-$Z$ few-electron atoms}, Physics Reports, 356(2002), pp. 119-228.
\bibitem{SHAB} V. M. Shabaev, I. I. Tupitsyn, V. A. Yerokhin, G. Plunien, and G. Soff, \emph{Dual kinetic balance approach to basis-set expansions for the Dirac equation}, Phys. Rev. Lett., 93(2004).
\bibitem{THA} B. Thaller, \emph{The Dirac Equation}, Springer-Verlag, Berlin, 1993.
\bibitem{TUPS} I. I. Tupitsyn and V. M. Shabaev, \emph{Spurious states of the Dirac equation in a finite basis set}, Optika i Spektroskopiya, 105(2008), pp. 203-209.
\bibitem{YOU} Y. You, J. S. Chen, and H. Lu \emph{Filters, reproducing kernel, and adaptive meshfree method}, Comput. Mech., 31(2003), pp. 316-326.
\bibitem{ZHAN} X. Zhang, X. Liu, M. Lu, and Y. Chen, \emph{Imposition of essential boundary conditions by displacement constraint equations in meshless methods}, Commun. Numer. Meth. Engng., 17(2001), pp. 165-178.
\bibitem{ZHA} S. Zhao, \emph{On the spurious solutions in the high-order finite difference methods for eigenvalue problems}, Comput. Methods Appl. Mech. Engng., 196(2007), pp. 5031-5046.
\bibitem{ZUP} C. Zuppa, \emph{Modified Taylor reproducing formulas and $h$-$p$ clouds}, Math. Comput., 77(2008), pp. 243-264.
\end{thebibliography}
\end{document}